\documentclass[12pt,twoside]{article}
\usepackage[utf8]{inputenc}
\usepackage{amsmath,graphicx,microtype,amssymb,setspace,verbatim,multirow,url,bbm,rotating,booktabs,amsthm}
\usepackage[font={normal}]{caption,subfig}
\usepackage{libertine}
\usepackage{color} 
\usepackage{sgame}
\usepackage{enumitem}
\usepackage{comment}
\usepackage{tikz}
\usetikzlibrary{positioning}
\usetikzlibrary{arrows.meta}
\usetikzlibrary{fit}

\newcommand{\R}{\mathbb R}

\newcommand{\E}{\mathbb E}

\newcommand{\Prob}{\mathbb P}

\makeatletter
	\renewcommand{\abstract}[1]{\def \@abstract {#1}}
	\newcommand{\jelcodes}[1]{\def \@jelcodes {#1}}
	\newcommand{\keywords}[1]{\def \@keywords {#1}}
	\newcommand{\thanknotes}[1]{\def \@thanknotes {#1}}
	\newcommand{\contact}[1]{\def \@contact {#1}}
	\newcommand{\shortauthor}[1]{\def \@shortauthor {#1}}
	\newcommand{\shorttitle}[1]{\def \@shorttitle {#1}}
\makeatother

\jelcodes{}
\thanknotes{}
\shortauthor{}
\shorttitle{}
\abstract{}

\usepackage[paperwidth=8.5in, paperheight=11in]{geometry}
\usepackage{setspace}
\usepackage[hang,flushmargin]{footmisc} 
\usepackage[authoryear]{natbib}

\makeatletter
\usepackage{fancyhdr}
\makeatother

\newcommand\blfootnote[1]{%
  \begingroup
  \renewcommand\thefootnote{}\footnote{#1}%
  \addtocounter{footnote}{-1}%
  \endgroup
}

\usepackage{authblk}
\makeatletter
\def \maketitle { 
	\thispagestyle{empty}
	\vspace*{0.1in}
	\blfootnote{\textsc{Contact.} \@contact.  \@thanknotes\\}
	\begin{center}
	\begin{minipage}{5.2in}
	\begin{center}
	{\large {\textbf{\@title}}}
	
	\vspace{0.2in}
	
	{\textsc{\@author}}
	
	\vspace{0.2in}
	
	{\@date}
	\end{center}
	
	\ifx\@abstract\@empty
	\relax
	\else
	{\small{\textsc{Abstract.} \@abstract}}
	\fi
	
	\ifx\@keywords\@empty
	\relax
	\else
	\vspace{0.2in}
	
	{\small\textsc{Keywords.} \@keywords.}
	\fi
	
	\ifx\@jelcodes\@empty
	\relax
	\else
	{\small\textsc{JEL Codes.} \@jelcodes.}
	\fi
	
	\end{minipage}
	\end{center} }
\makeatother

\makeatletter
\def\@seccntformat#1{\csname the#1\endcsname.\ }
\makeatother

\usepackage{sectsty}
\allsectionsfont{\noindent\normalsize}
\sectionfont{\centering\normalsize\uppercase}
\paragraphfont{\textnormal}

\providecommand{\U}[1]{\protect\rule{.1in}{.1in}}
\newtheorem{theorem}{Theorem}

\newtheorem{corollary}{Corollary}

\newtheorem{definition}{Definition}
\newtheorem{example}{Example}

\newtheorem{lemma}{Lemma}

\newtheorem{proposition}{Proposition}

\usepackage{hyperref}

\makeatletter
\renewcommand\@biblabel[1]{}
\makeatother

\begin{document}

\title{\Large Sequentially Optimal Pricing under Informational Robustness} 
\author{\textsc{Zihao Li* ~~~~~ Jonathan Libgober** ~~~~~ Xiaosheng Mu***}}

\affil{*Columbia University \\ **University of Southern California \\ ***Princeton University}

\date{\today}

\abstract{A seller sells an object over time but is uncertain how the buyer learns their willingness-to-pay. We consider informational robustness under \textit{limited commitment}, where the seller offers a price \textit{each period} to maximize expected continuation profit against worst-case learning. Our formulation considers the worst case \textit{sequentially}. We characterize an essentially unique equilibrium under general conditions. We further show that, under mild conditions on the prior distribution, the equilibrium profit coincides exactly with the profit guaranteed by the equilibrium price path even under arbitrary (unrestricted) learning processes.
}


\shortauthor{Li, Libgober and Mu}
\shorttitle{Robust Pricing without Commitment}

\contact{\textsc{zl3366@columbia.edu, libgober@usc.edu, xmu@princeton.edu}}
\thanknotes{We thank Dirk Bergemann, Alex Bloedel, Daniel Clark, Drew Fudenberg, Johannes H\"orner, Navin Kartik, Deniz Kattwinkel, Peter Klibanoff, George Mailath, Rob Metcalfe, Mathias Nunez, Alessandro Pavan, Jo\~ao Ramos, Andy Skrzypacz, Bal\'azs Szentes and Kai Hao Yang for helpful comments and seminar audiences at Boston University, Brown, Harvard/MIT, LSE, NASMES, Rochester, SITE: Dynamic Games, Stony Brook, UChicago, UPenn, USC and Yale for feedback. The second author thanks the hospitality of the Cowles Foundation and Yale University, which hosted him during part of this research.}

%Nageeb Ali, Sarah Auster, Alex Bloedel, Gabriel Carroll, Daniel Clark, Songzi Du, Andrew Ellis, Jack Fanning, Mira Frick, Drew Fudenberg, Kevin He, Johannes H\"orner, Ryota Iijima, Yuhta Ishii, Navin Kartik, Andreas Kleiner, Giacomo Lanzani, Bart Lipman, Andrew McClellan, Francesco Nava, Jo\~ao Ramos, Doron Ravid, Joseph Root, Larry Samuelson, Andy Skrzypacz, Bal\'azs Szentes, 
\setcounter{page}{0}

\maketitle

\newpage

\section{Introduction} We consider a seller who aims to sell one unit of an indivisible good to a buyer, but faces \emph{limited commitment}, only able to commit to a price within the period it is offered.  The buyer, in turn, may not initially know her willingness-to-pay, but can learn it through some (potentially complex) process while considering purchase. We propose a new approach, motivated by the literature on informational robustness, to analyze these settings when the seller reoptimizes each period to maximize payoffs \emph{from that time onward}. 

A buyer may lack precise knowledge of her willingness-to-pay for many reasons. Consider, for example, a new parent negotiating to buy a house from a previous homeowner who has already vacated.\footnote{Such negotiations often lack mechanisms to avoid price revisions. The sale of land is the leading example in \cite{Coase}. \cite{HS2015} discuss recent evidence on post-match price revisions in housing markets.} Neighborhood characteristics—such as average annual NOx levels, risks from natural disasters, or school quality—may not be immediately apparent.\footnote{Information availability about idiosyncratic value-relevant characteristics is particularly well-documented in housing markets. \cite{RobsPaper} conduct an experiment showing that providing information about flood risk significantly changes buyer behavior. \cite{AinsworthEtAl} show that households often lack accurate information about local school quality, and that belief accuracy predicts child achievement. \cite{BergmanEtAl} show experimentally that providing school information influences household location choices.} A first-time buyer may not even initially understand how such characteristics \emph{should} translate into a true willingness-to-pay. 

Our starting point is the observation that buyer learning can take many forms in such settings. The buyer might be perfectly informed at the outset---or not. She might consult family or various online resources, or even learn about when further information will arrive; an ``unexpected'' offer could influence the sources she considers. This feature is important because learning can influence purchase timing---for example, by incentivizing delay to obtain more information. Given the wide range of possibilities, imposing any specific structure on buyer learning risks misrepresenting the actual problem of interest. Such richness in the set of possible learning processes is a fundamental feature of many economically significant interactions.

In principle, the possibility of learning could fundamentally alter behavior in applications where the seller lacks commitment to future prices. This follows from the standard intuition that the degree to which delay selects low willingness-to-pay buyers shapes equilibrium pricing (see Section \ref{sect:literature} for discussion). While incentives to delay under a given price path are determined by expected surplus absent learning, this logic can break down once learning is possible. A buyer who expects to value the product highly might nevertheless delay to become more certain that this is the case. While the economic significance of limited commitment is well recognized, relatively little work has addressed its interaction with unrestricted buyer learning.\footnote{As we discuss in Section \ref{sect:literature}, existing precedents impose particular structure on learning, generally excluding many of the possibilities outlined here. As \citet{PavanSurvey} notes: ``The literature on limited commitment has made important progress in recent years.... However, this literature assumes information is static, thus abstracting from the questions at the heart of the dynamic mechanism design literature. I expect interesting new developments to come out from combining the two literatures.'' We view our contribution in the spirit of this agenda.}

In fact, without restrictions on learning, constant price paths can be sustained and multiple equilibria can arise—a stark departure from outcomes without learning (see Section \ref{sect:positiveselection}). We view this as a proof of concept that some structure is necessary to derive interpretable pricing heuristics or to characterize equilibrium properties. Yet in many cases such restrictions lack economic justification, motivating a different perspective to serve as a useful baseline.

Our approach focuses on situations in which the seller is completely ignorant about how the buyer learns. Even if an inexperienced seller were able to form a prior over the buyer's true willingness-to-pay using market data (e.g., transaction prices), he may nevertheless know nothing about her information sources. A seller unfamiliar with up-to-date resources popular among buyers may lack the confidence needed to form a prior over the set of learning processes, and thus may not anticipate any \emph{particular} one. Assuming the confidence needed to form such a prior is often unrealistic as the information in question is typically intangible, making it unclear how a seller would be able to specify such a concrete belief, especially when it may evolve over time.

This motivation---to study the implications of uncertainty over buyer information---is also central to work in the literature on informational robustness \citep{BBM2017,FirstBrooksDu,BrooksDuDuality,DR2023,koh2025robust}.\footnote{One reason for the popularity of the \emph{informationally} robust approach in particular is the influence of the \emph{Wilson Critique}, which emphasizes that strong epistemic assumptions in mechanism design severely limit its applicability.} The standard formalization of “good performance under complete ignorance” associates this goal with a favorable worst-case guarantee across the set of possibilities considered. This formulation is also common when uncertainty is unrelated to information per se \citep{BM2005,Carroll_2015,Carroll_2017,LRS2020,CZ2022,HL2022}. 

Despite this interest, there is no consensus on how to \emph{formally} define worst-case optimality in dynamic settings without commitment. A key difficulty is the potential inconsistency of worst-case solutions over time.\footnote{See \citet{CarrollSurvey} for discussion of this issue.} To illustrate, suppose that at time 2 the seller fears that the buyer will perfectly learn her value at time 10, encouraging a “wait-and-see” strategy. Yet once time 10 arrives, such learning might not be the worst for the seller; keeping a buyer with a value slightly above the price ignorant could deter purchase. The worst-case objective at time 2 would thus conflict with that at time 10. Seeking favorable worst-case performance across time can thus lead to contradictory conjectures about what this entails, with no clear principle reconciling them.

This gap in the literature is significant: it makes dynamic models more dependent on correctly specified Bayesian beliefs about the informational environment than static models, even though this assumption may be \emph{more} demanding under dynamics. Sometimes the reason a seller lacks confidence to form a precise prior (e.g., limited experience selling one's house) also suggests a lack of commitment. We aim to provide a concrete step toward the agenda described by \citet{BergemannValimakiSurvey}, writing that dynamic mechanism design has focused on “... Bayesian solutions and relied on a shared and common prior of all participating players. Yet, this clearly is a strong assumption and a natural question would be to what extent weaker informational assumptions, and corresponding solution concepts, could provide new insights into the format of dynamic mechanisms.” While our approach still involves strong assumptions, offering \emph{some} alternative that makes the conceptual issues concrete is, we believe, a natural step toward fulfilling this agenda.

\subsection{Our Approach}

We consider a protocol in which (i) the seller posts prices over time and (ii) the buyer learns about her willingness-to-pay. In each period, the seller offers the object at a chosen price, and the buyer---after receiving her current information---either accepts or rejects the offer. The interaction continues until a terminal date or indefinitely. This environment reduces to a standard model if the buyer knows her willingness-to-pay from the outset; under a Bayesian approach, the seller would then form a prior over the buyer's learning process.

We propose a robust objective in which, at every period, the seller assumes that the buyer's information structure \emph{minimizes the seller’s continuation profit from that period onward}. We call such a process \emph{sequentially worst-case}. This benchmark preserves \emph{dynamic consistency}: the worst case the seller anticipates for tomorrow coincides with the ``realized'' worst case once tomorrow arrives---both on and off the equilibrium path. This strikes us as a natural starting point for extending the robust approach to settings without commitment, given the conceptual issues that arise otherwise. Dynamic consistency in maxmin models has been extensively studied in decision theory, and whether it is \emph{per se} desirable remains a matter of debate.\footnote{For example, \cite{AlNajjarWeinstein} document behavioral anomalies that can emerge in dynamic maxmin models without dynamic consistency, while \cite{Siniscalchi} argues that several of these anomalies may be natural. See also \cite{ES_2003} for discussion of why dynamic consistency may be desirable. Related inconsistencies can arise in sequential games with ambiguity even outside of worst-case information; see \cite{BCLM2019,BFLM2019} for versions under \emph{smooth-ambiguity preferences}, which approximate maxmin in a limit.}

While we do not engage further in this debate, our setting is fundamentally different from single-agent applications of such formulations \citep{ACM2022,Malladi2023}. A defining feature of robust mechanism design is the presence of another strategic agent; here, the buyer may follow any behavior consistent with rationality.

Following common practice in the robustness literature, we formalize worst-case information via an “adversarial Nature.” In each period, Nature chooses the buyer’s information for \emph{that} period to minimize discounted continuation profit. The worst-case learning process then emerges as an equilibrium object. Viewing Nature’s choices as part of equilibrium is a shortcut for pinning down the seller’s conjectures about the learning process both on and off path. Doing so amounts to ensuring that: (i) in every period the conjecture minimizes seller profit, and (ii) the realized outcomes are consistent with earlier conjectures about what those outcomes would be. The first property explains why the benchmark is worst-case, and the second why dynamic consistency holds—when the time comes for an information structure to be the worst case, it is. 

Presenting the model using equilibrium notions allows us to leverage traditional intuitions and techniques in formally analyzing dynamic worst-case solutions. Section \ref{sect:saddle} presents an alternative formulation which shows that the substantive assumption is rather that the solution concept rules out certain ways of coordinating buyer behavior with future information.\footnote{Formally, Section \ref{sect:saddle} shows this setup admits an interpretation as a saddle-point under a particular objective, similar to other (static) robust mechanism design exercises. While that objective does not require a literal interpretation of Nature, we believe directly modeling the buyer–seller interaction as a dynamic game is more transparent.} Thus, our exercise does not require interpreting Nature literally, even though describing the model as we do yields a convenient and transparent framework for analysis.

\subsection{Our Results}

Our main observations are the following twin results:

\begin{enumerate}
\item \textbf{Equilibrium characterization.} We characterize equilibrium outcomes and show that the equilibrium is essentially unique,\footnote{In the sense of \citet{FLT1985}: the seller’s and buyer’s actions are deterministic after possible seller randomization in the first period.} with the following property: in any period, the buyer’s decision is identical to the decision she would make if she considered only the information available in that period, as though no further learning would occur. This property rules out the dramatic departures from standard predictions that can arise under Bayesian approaches.

\item \textbf{Safety condition.} We provide a permissive condition under which the seller’s equilibrium profit coincides exactly with the profit guaranteed by the equilibrium price path across \emph{all} learning possibilities. In such cases, restricting attention to sequentially worst-case learning is not restrictive at all.
\end{enumerate}

\noindent Our equilibrium characterization shows that sequentially worst-case learning minimizes the probability of sale \emph{within each period} given the price path. While our model assumes the worst case is chosen to minimize \emph{total discounted profit}, one might expect that a buyer anticipating future information could be deterred from buying in the present. We show this does not occur under sequentially worst-case learning. This reduction allows us to characterize equilibrium explicitly and to recover the familiar structure of the form of equilibrium price paths. In particular, the seller does not randomize on the equilibrium path, except possibly in the initial period.

The intuition for why the worst case minimizes the within-period sale probability comes from backward induction. In the final period, worst-case information has the property that any buyer who does not purchase is indifferent between purchasing and not, making her payoff the same regardless of her action. In particular, her payoff would be unchanged were she to instead \emph{always} purchase, despite her breaking indifference by not purchasing. Hence, in the second-to-last period, the buyer behaves as if no future information will arrive. The worst-case information in that period thus simply minimizes the probability of purchase, and the same logic propagates backward.

Our second main result shows that, for a large class of environments, sequentially worst-case learning minimizes the seller’s profit across \emph{all} learning processes, \emph{fixing} the pricing strategy to be the one that arises in equilibrium. This preserves the normative appeal of the worst-case objective---good performance against unrestricted learning---while ensuring tractability. Put differently, even if the seller were misspecified about the buyer’s learning process (e.g., if the dynamic consistency constraints need not hold), this misspecification could not reduce her payoff. We call any equilibrium pricing strategy with this property a \emph{safe solution}. 

This result is conceptually subtle because it requires the worst case to be defined \emph{holding fixed} the seller’s pricing strategy. This differs from the alternative approach in which one considers the worst-case equilibrium between seller and buyer that can emerge given an arbitrary learning process. It is possible that some other pricing strategy could arise in equilibrium for some (non–sequentially worst-case) learning process, for which the seller would be worse off.\footnote{Indeed, under fairly general conditions, one can construct learning processes and corresponding equilibria in which the seller earns less than under the sequentially worst-case criterion. But subjectively, these outcomes do not strike us as sharing the intuitive appeal of sequentially worst case learning.} Intuitively, providing more information in later periods involves a tradeoff, between: (i) inducing more delay by higher-value buyers in earlier periods, and (ii) inducing more---or at least earlier---sales by lower-value buyers in later periods. The condition on the prior value distribution that we identify ensuring the price path is a safe solution---which we call \emph{threshold-ratio monotonicity}---essentially requires that the increase in willingness-to-delay from additional future information is small relative to the additional sales such information generates. While this property does rule out some cases, it holds for many distributions considered in past work \citep[e.g.,][]{FS2013}.

\subsection{Our Message}

Our work achieves twin goals. First, we provide a baseline for understanding how the possibility of learning interacts with limited commitment to prices. Despite several subtleties we identify, the benchmark recovers traditional intuition and heuristics: the seller lowers prices over time to sell to the residual (lower-value) buyers. We view this formulation as a useful step toward studying outcomes under limited commitment without imposing restrictive structure on learning---especially since (a) Bayesian approaches have not yet provided a clear benchmark for the combination of general learning and limited commitment, and (b) insofar as they could, our analysis suggests they may instead imply ``anything goes,'' making clear insights more difficult to obtain.

The second goal is to enrich the informationally robust approach to address conceptual difficulties related to limited commitment. The literature on informational robustness often points to the challenges in determining practically relevant learning processes as an economic motivation. Despite dynamic consistency issues, we see no reason such concerns would be irrelevant absent commitment. Our result that the equilibrium price path is a safe solution suggests that dynamic consistency concerns may be less restrictive than previously thought. Our framework thus provides a template for interrogating such concerns formally. 
%We hope our work inspires others to seek to move beyond stringent Bayesian approaches when alternatives may yield sharper predictions and simpler intuitions.  

\subsection{Relevant Literature}  \label{sect:literature}

The literature on robust mechanism design originated from the goal of relaxing the strong common-knowledge assumptions inherent in Bayesian mechanism design \citep{BM2005}. Early work assumed agents knew their own preferences; later work on \emph{informationally} robust mechanism design (e.g., \cite{Du2018}) allowed designers to be uncertain about what agents know about their own preferences. To our knowledge, relatively few papers have studied informationally robust mechanism design in dynamic settings, and issues related to commitment are rarely addressed. An exception is \cite{koh2025robust}, although in that setting the commitment solution is implementable without commitment—a stark contrast to the dynamic durable goods sales problem here.\footnote{\cite{firstpaper} studied robust dynamic pricing with commitment, avoiding the consistency and equilibrium characterization challenges we face. Other dynamic robust mechanism design papers include \cite{chassang2013calibrated,penta2015robust,durandard2024robust}, but in these and most others, the worst case is considered only once.} Limited commitment fundamentally requires evaluating the worst case repeatedly, a defining feature of our exercise.

If the buyer knew her realized willingness-to-pay, our model would reduce to the textbook durable goods monopoly without commitment \citep{FLT1985,GSW,AusubelDeneckere1989}. While we use some technical results from this literature, our focus is largely orthogonal: the $\delta \rightarrow 1$ limit is not our main interest, unlike most papers in this area. A notable exception is \cite{FS2013}, which characterizes \emph{non-trivial} pricing dynamics with “frequent offers” and a finite time to trade.

Changes in preferences can alter the conclusions of the Coase conjecture literature (e.g., \cite{Ortner2017,Ortner2022,AcharyaOrtner2017}), and learning can be interpreted as a form of preference change. Relatedly, \cite{Lomys2018}, \cite{Duraj2020}, and \cite{LaihoSalmi2020} study how Coasian dynamics are affected by buyer learning, but under restrictive assumptions on type distributions or learning processes. These effects arise from the interaction between learning and selection; the importance of the direction of selection for Coasian dynamics is highlighted in \cite{Tirole2016} and \cite{AliKartikKleiner2023}. In contrast, one interpretation of our main result is that Coasian dynamics are fully restored under our informationally robust objective. In Appendix \ref{sect:alternatives}, we discuss alternative dynamic worst-case formulations under which the prospect of learning introduces additional forces, breaking this restoration.

A less directly related literature considers mechanism design where agents—rather than the designer—have non-Bayesian preferences, including the maxmin case \citep{BoseRenou2014,Wolitzky2016,DiTillioEtAl2017}. The typical focus there is on how the \emph{designer} can exploit such preferences; some papers explicitly examine exploiting dynamic inconsistency \citep{BoseEtAl2006,BoseDaripa2009}.

Finally, our \emph{safe solution} requirement connects to recent proposals to strengthen robustness criteria—seeking not only optimality against a \emph{single} worst case, but also good performance across a broader set of possibilities. \cite{kambhampati2025proper} pursues this goal, but evaluates performance against alternative possibilities from the \emph{same} set as the one initially used to define worst-case performance. \cite{ball2025robust} also considers mechanism performance under ambiguity set expansions, as the safe solution requirement does, but focuses on local expansions---requiring near-optimal performance when Nature’s choices are ``close'' to those the designer entertains. By contrast, the safe solution concept here considers \emph{arbitrary} alternative possibilities.

\section{Model} \label{sect:model}
We first present the basic primitives of the environment. We then turn to the mechanics of how the buyer and seller interact, describing how strategies and beliefs are defined as well as how learning works in our model. The timing of moves within each period in the game is summarized in Figure \ref{fig:timing_multi}. Section~\ref{subsect:equilibrium} introduces our worst-case notion. To maintain focus, discussion of model assumptions is deferred to Section~\ref{subsect:discussion}, and alternative worst-case notions are covered in Appendix~\ref{sect:alternatives}.
\subsection{Environment} \label{subsect:underlying}
A seller (he) of a durable good (e.g., a house) interacts with a single buyer (she) in discrete time until a terminal date $T \leq \infty$.\footnote{We handle the cases $T = \infty$ and $T < \infty$ separately.}  
The buyer can purchase the good at any time $t = 1, \ldots, T$. She has unit demand and obtains utility $v$ from purchasing, where $v$ is drawn once at time $0$ from a commonly known distribution $F$ and remains fixed throughout the game.  
The distribution $F$ has density $f$ supported on a compact interval $[\underline{v}, \overline{v}] \subset \mathbb{R}_{+}$. The case $\underline{v} > 0$ is referred to as the ``gap case,'' while $\underline{v} = 0$ is the ``no-gap case.''  
Both buyer and seller discount payoffs by a common factor $\delta \in [0,1)$.

However, neither the buyer nor the seller observes the realization of $v$. Instead, the buyer may learn about $v$ over time. Formally, an \emph{information structure} is a pair $(S,\pi)$, where $S$ is a standard Borel space and $\pi : [\underline{v}, \overline{v}] \to \Delta(S)$ is a measurable mapping from valuations to probability distributions over signals. In the dynamic setting, the buyer receives signals according to such information structures in a history--dependent manner, as described below.

%For now, observe that the buyer can form her posterior following $(I_{1}, \ldots, I_{t})$ and $(s_{1}, \ldots, s_{t})$ as:
%\vspace{-4mm}
%\begin{equation} 
%v \mid I_{1}, s_{1}, \ldots, I_{t}, s_{t}, \label{eq:BayesUpdating}
%\end{equation}

%\noindent via Bayesian updating (and with no other information). 

\subsection{Actions and Histories} \label{subsect:timing}

Each period $t$ begins with the seller choosing a price $p_{t} \in \mathbb{R}_{+}$. While the seller may randomize over prices, we assume that the realization of $p_{t}$ is observed before the game continues. 

After the price $p_{t}$ is realized, the buyer observes a signal drawn according to some information structure, determined in that period. We use the expositional device, common in the robustness literature, that this information structure is chosen by some player, referred to as ``Nature.'' Later in the paper, we will clarify in more depth what this expositional device delivers substantively, but for now we simply mention that this simplifies our presentation of our solution concept. Throughout the paper, we assume that the information structure as well as the realized signals are observed only by the buyer, and \emph{not} by the seller. 

After observing the price for the given period as well as the new information described in the previous paragraph, the buyer then updates beliefs and decides whether to purchase or not. Formally, the buyer’s action in period $t$ is denoted $a_{t} \in \{0,1\}$, where $a_{t} = 1$ indicates purchase at price $p_{t}$ and $a_{t} = 0$ indicates no purchase. If $a_{t} = 1$ or $t = T$, the game ends; otherwise, play proceeds to period $t+1$.

Notice that the information sets for each player---the seller, Nature, and the buyer---are distinct. Since the buyer’s only decision is whether to purchase, and since the game ends once she does so, our definition of histories will condition on the event that the buyer has not yet purchased. We let $H_{S}^{t}$, $H_{N}^{t}$, and $H_{B}^{t}$, denote the set of possible information sets for the seller, Nature, and the buyer, respectively, at time $t$. The set of seller histories, $H_{S}^{t}$, is equal to the set of possible price histories before time $t$. The set of Nature histories, $H_{N}^{t}$, is the set of price histories up to and including time $t$, together with the set of all possible information structures and signal realizations before time $t$. Finally, $H_{B}^{t}$ is equal to the same set of sequences as $H_{N}^{t}$, concatenated with the information structure and signal realization for period $t$. Let $\mathcal{H}$ denote the set of all such histories.

Although we have not yet specified how learning is determined, it is worth noting that the framework so far is standard. For example, the environment reduces to textbook bargaining with one-sided private information if $\pi_{1}$ revealed $v$ to the buyer. The novelty lies in allowing the buyer to acquire information about $v$ over time, and in specifying a solution concept which allows for endogenous interactions between this information and the seller's strategy.

\subsection{Defining Strategies and Beliefs} \label{subsect:strategies}

We now describe strategies and beliefs for the players. This step helps us define our solution concept capturing the seller's informationally robust objective. As discussed in the introduction, part of our contribution lies in formulating such an objective given the seller's limited commitment. To our knowledge, there is no consensus approach on how to do so. 

\paragraph{\textbf{Strategies.}}  A \emph{pricing strategy} is a function
\[
\sigma : \bigcup_{t} H_{S}^{t} \to \Delta(\mathbb{R}_{+}),
\]
so that for each seller information set, $\sigma$ specifies a distribution over prices. A \emph{price path} is a sequence $(p_{1},p_{2},\ldots)$; we write $p^{t} = (p_{1},\ldots,p_{t})$ for the history of prices up to period $t$. A \emph{learning process} is a function
\[
\Pi : \bigcup_{t} H_{N}^{t} \to \Delta\left(\left\{(\pi,S)\right\}\right),
\]
which assigns to each of Nature’s information sets a distribution over the signal space and the information structure to be used in that period. A \emph{buyer strategy} is a function 
\[
\alpha : \bigcup_{t} H_{B}^{t} \to \Delta(\{0,1\}),
\]
where, for each buyer information set, $\alpha$ specifies a probability distribution over $\{0,1\}$: as mentioned, 
$0$ denotes ``not buying,'' and $1$ denotes ``buying.''\footnote{In the standard Coasian bargaining model with a continuum of types, mixed strategies for the buyer are unnecessary for equilibrium existence. However, with general learning possibilities, mixed strategies may be necessary. }

\paragraph{\textbf{Beliefs.}}  
Two histories $h$ and $h'$ are \emph{non-contradictory} if they coincide whenever possible for them to do so.  
Given $(\sigma,\alpha,\Pi)$ as above, define
\[
\mathbb{P}_{\sigma,\alpha,\Pi} : \mathcal{H} \to \Delta([\underline{v},\overline{v}] \times \mathcal{H})
\]
to be the probability distribution over $(v,h')$ induced when starting from history $h \in \mathcal{H}$ and given the strategy profile $(\sigma,\alpha,\Pi)$. For every $h \in \mathcal{H}$, $\mathbb{P}_{\sigma,\alpha,\Pi}(h)$ is supported on histories $h'$ that are non-contradictory with $h$.

A \emph{belief system} for player $j \in \{S,B,N\}$ is a function
\[
\mu_{j} : \bigcup_{t} H_{j}^{t} \to \Delta([\underline{v},\overline{v}] \times \mathcal{H}),
\]
such that at each of $j$’s information sets, $\mu_{j}$ is supported on histories $h \in \mathcal{H}$ consistent with that information set. Let $\mu = (\mu_{S},\mu_{B},\mu_{N})$.

A belief system \emph{satisfies Bayes’ rule where possible} if, for $t < s$ and $h^{t}$ non-contradictory with $h^{s}$, $\mu(h^{s})$ is obtained from $\mu(h^{t})$ via Bayes’ rule. Since Nature and the buyer observe the complete history of prices, information structures and signal realizations, their beliefs concern only the private type $v$. By contrast, the seller observes neither the information structure nor the signal realization; he observes only past prices. Accordingly, the seller’s beliefs are defined over both $v$ and the unobserved history of information structures and signals. For any given seller information set, these beliefs are degenerate on the observed prices.

\begin{figure}[t!]
\centering
\resizebox{1\textwidth}{!}{%
\begin{tikzpicture}[
    node distance=1.2cm and 1.5cm,
    every node/.style={font=\small},
    >=stealth
]

% Period t label
% \node (period) [font=\bfseries] {Period $t$};

% Seller
\node (seller) [draw, rectangle, minimum width=3cm,text width=3cm, minimum height=0.8cm, below=0.3cm,  align=center] 
    {\textbf{Seller chooses price $p_t$}: \\ only observes price history $p_{1}, \ldots, p_{t-1}$ };
%\node (sellerInfo) [below=0.10cm of seller, font=\scriptsize, align=center] 
%    {Observes $(p_1,\ldots,p_{t-1})$ only\\Does \emph{not} observe buyer's signals};

\node (dotsLeft) [left=.75cm of seller, font=\large] {$\cdots$};
\draw[->, thick] (dotsLeft.east) -- (seller.west);

% Nature
\node (nature) [draw, rectangle, minimum width=3.5cm,text width=3.5cm, minimum height=0.8cm, right=of seller, align=center] 
    {\textbf{Nature chooses information $\pi_{t}$}: Observes both price history $p_{1}, \ldots, p_{t}$ and  $(\pi_{1}, s_{1}),$ $\ldots, (\pi_{t-1}, s_{t-1})$};
%\node (natureInfo) [below=0.10cm of nature, font=\scriptsize, align=center] 
 %   {Observes complete history};

% Buyer
\node (buyer) [draw, rectangle, minimum width=4.2cm, minimum height=0.8cm, right=of nature, align=center, text width=4.2cm] 
    {\textbf{Buyer updates beliefs}, \textbf{then chooses} to buy ($a_{t}=1$) or not ($a_{t}=0$) at price $p_{t}$};
%\node (buyerInfo) [below=0.10cm of buyer, font=\scriptsize, align=center] 
 %   {Observes complete history};

% NEW: Continuation box (to the right of Buyer)
%\node (continue) [draw, rectangle, minimum width=1.6cm, minimum height=0.8cm, right=of buyer, align=center, text width=1.6cm] 
%    {Game continues};
%\draw[->, thick] (buyer) -- (continue);

\node (dotsRight) [right=2.3cm of buyer, font=\large] {$\cdots$};
\draw[->, thick] (buyer) -- (dotsRight)
    node[midway, sloped, above] {$a_t=0$ and}
    node[midway, sloped, below]{$t<T$};

% End condition (now in a box, below Buyer)
\node (endCond) [draw, rectangle, minimum width=1.5cm, minimum height=1cm, below=1cm of buyer, align=center, text width=1.5cm] 
    {Payoffs realized};
\draw[->, thick] (buyer) -- (endCond)
     node[midway, align=center,left,xshift=.4mm] {$a_t=1$}
     node[midway, align=center,right,xshift=-.4mm,yshift=.35mm] {or $t=T$};

% Loop back arrow for continuation (continue → seller)
%\draw[->, thick] (continue.south) |- (seller.north);

% Arrows for within-period moves
\draw[->, thick] (seller) -- (nature);
\draw[->, thick] (nature) -- (buyer);

% End or continue decision
%\node (endCond) [below=1.2cm of buyer, font=\small, align=center] 
%    {If $a_t=1$ or $t = T$,\\ game ends};

%\draw[->, thick] (buyer) -- (endCond);

% Loop back arrow for continuation
%\node (continue) [left=5.0cm of endCond, font=\small, align=center] 
%    {If $a_t=0$ and $t<T$,\\ proceed to $t+1$};
%\draw[->, thick] (endCond.west) -- ++(-0.8,0) |- (seller.south);

% Time flow label
%\node[above=0.3cm of nature, font=\small\itshape] 
%   {Within period: Seller $\to$ Nature $\to$ Buyer};

\node[draw, thick, rounded corners, inner sep=5pt, fit=(dotsLeft)(dotsRight)(endCond)(nature)] {};

\end{tikzpicture}%
}
\caption{Timing of moves and information sets within some period $t$.}
\label{fig:timing_multi}
\end{figure}

\subsection{Equilibrium Assumptions} \label{subsect:equilibrium}

We finally specify our solution concept. Fix an arbitrary triple $(\sigma,\alpha,\Pi)$, the induced measure $\mathbb{P}_{\sigma,\alpha,\Pi}$, and a belief system $\mu$. Let $h_{S}^{t}, h_{B}^{t}, h_{N}^{t}$ denote representative decision nodes for the seller, buyer, and Nature, respectively, at time $t$. Let $F_{i}^{t}(\cdot)$ denote the prior $F(\cdot)$ conditional on $h_{i}^{t}$ for $i \in \{S,B,N\}$.

The buyer’s strategy is \emph{sequentially rational} given $\mu$ if, for all $h_{B}^{t}$, the action prescribed by $\alpha$ maximizes the buyer’s expected continuation payoff conditional on reaching $h_{B}^{t}$:
\[
\mathbb{E}_{\mu,\mathbb{P}_{\sigma,\alpha,\Pi}}\!\left[\sum_{\tau \ge t} \delta^{\tau-t} (v-p_{\tau})\,\mathbf{1}_{\{\text{accept at }\tau\}} \,\big|\, h_{B}^{t}\right],
\]
where $\tau$ is the induced stopping time.

If the buyer purchases at some time $s$ at price $p_{s}$, then from the perspective of time $t<s$ the seller obtains payoff $\delta^{\,s-t} p_{s}$. The seller’s pricing strategy is \emph{sequentially rational} given $\mu$ if, at every $h_{S}^{t}$, the action prescribed by $\sigma$ maximizes the seller’s expected continuation payoff conditional on reaching $h_{S}^{t}$:
\begin{equation}
\mathbb{E}_{\mu,\mathbb{P}_{\sigma,\alpha,\Pi}}\!\left[\sum_{k=t}^{T} \delta^{k-t} p_{k}\,\mathbf{1}_{\{a_{k}=1\}} \,\big|\, h_{S}^{t}\right],
\label{eq:sellerpayoff}
\end{equation}
where $a_{k}\in\{0,1\}$ denotes the buyer’s acceptance decision at time $k$.

The innovation behind our solution concept stems most directly from our specification of how the learning process is realized, which is as follows: We say that a learning process is \emph{sequentially worst case given $\mu$} if, at every $h_{N}^{t}$, Nature’s action prescribed by $\Pi$ \emph{minimizes} the seller’s expected continuation payoff conditional on reaching $h_{N}^{t}$:
\begin{equation}
\mathbb{E}_{\mu,\mathbb{P}_{\sigma,\alpha,\Pi}}\!\left[\sum_{k=t}^{T} \delta^{k-t} p_{k}\,\mathbf{1}_{\{a_{k}=1\}} \,\big|\, h_{N}^{t}\right].
\end{equation}

\begin{definition}\label{def:wctcc}
Let $(\sigma,\alpha,\Pi)$ be strategies for the seller, buyer, and Nature, respectively, and let $\mu$ be a belief system. The quadruple $((\sigma,\alpha,\Pi),\mu)$ is an \textbf{equilibrium} if and only if:
\begin{itemize}
\item $\alpha$ is sequentially rational for the buyer;
\item $\sigma$ is sequentially rational for the seller;
\item $\Pi$ is sequentially worst case;
\item $\mu$ is derived from strategies using Bayes’ rule wherever possible; and
\item The distribution of $v$ induced by $\mu$ at time $t$ is \textbf{consistent} with Bayesian updating via information structures and signal realizations,
\begin{equation}
v \,\big|\, (\pi_{1},s_{1}),\ldots,(\pi_{t},s_{t}),
\label{eq:BayesUpdating}
\end{equation}
where $(\pi_{k},s_{k})$ denotes the information structure chosen and signal realized in period $k$.
\end{itemize}

\end{definition}

\noindent Notice that according to Definition \ref{def:wctcc}, the distribution of $v$ induced by $\mu$ at time $t$ does not depend on prices charged, signal spaces chosen, and so on. Thus, the buyer only updates beliefs about $v$ based on the learning process---crucially, on or off path.\footnote{This restriction is in the spirit of “no-signalling-what-you-don’t-know” refinements \citep{fudenberg1991perfect}. Otherwise, one could construct equilibria in which a deviation is deterred by the buyer adopting a belief that $v=0$ with probability 1, even if this lies outside the support of $F$.} Our goal in this paper is to determine equilibrium outcomes.

We emphasize that, although our exposition imposes equilibrium conditions on Nature, this should \emph{not} be read as attributing it any intrinsic motivation. The formulation serves only as an expositional device to capture how the seller can correctly anticipate worst-case outcomes both on-path and following deviations. In Section \ref{sect:saddle}, we make this point precise by showing that the solution admits an equivalent saddle-point interpretation, for which no such device is needed.

\section{Solution of the Main Model} \label{sect:solutionBaseline}

We now characterize both the equilibrium price path and the induced learning process under sequentially worst-case learning. Section~\ref{sect:safe} examines worst-case expected seller profit without restricting to learning processes that emerge as sequentially worst-case.

\subsection{Arbitrary Finite Horizon}

The sequentially worst-case learning process turns out to be characterized by a sequence of thresholds adapted to the seller's pricing strategy.

\begin{definition}
Fix any seller strategy $\sigma$. For any realized $p_{t}$, let $w_{t}(p_{t},\sigma)$ be the unique value satisfying
\begin{equation*}
w_{t}(p_{t},\sigma) - p_{t} \;=\; \max_{\tau \ge t+1} \mathbb{E}\!\left[\delta^{\,\tau-t}\big(w_{t}(p_{t},\sigma) - p_{\tau}\big) \,\middle|\, p^{t} \right],
\end{equation*}
where $\tau$ ranges over all stopping times.  

The \textbf{myopic threshold} learning process is defined as the learning process where, at any $h_{N}^{t}$, the buyer learns whether $v > y_{t}$, where $y_{t}$ satisfies
\begin{equation}
w_{t}(p_{t},\sigma) \;=\; \mathbb{E}_{v \sim F^{t}_{N}}\!\left[\, v \,\middle|\, v \le y_{t} \,\right].
\label{eq:fullymyopicthreshold}
\end{equation}
\end{definition}

The terminology reflects the property that $y_{t}$ minimizes the probability of purchase at time $t$, given the continuation strategy $\sigma$ when $v \sim F_{N}^{t}$. This problem, in turn, is equivalent to a static Bayesian persuasion problem in which Nature (as Sender) seeks to induce the buyer (as Receiver) not to purchase. While Nature’s objective is formally to minimize the seller’s \emph{discounted profit}, in our setting this coincides with minimizing the \emph{period-$t$ sale probability}.

\begin{theorem} \label{thm:baseline}
When $T<\infty$, the equilibrium price path is unique and deterministic (following possible randomization in period $t=1$) and weakly decreasing over time. The buyer behaves according to the myopic threshold learning process with respect to the equilibrium price path.\footnote{Nature may in principle provide more information than the threshold structure, provided this does not alter the buyer’s behavior.} 
\end{theorem}

\subsubsection{Illustrating Theorem \ref{thm:baseline} with $T=2$ and Uniform Values} \label{sect:twoperiods}

While the proof of Theorem~\ref{thm:baseline} is involved, most of the economic intuition can be seen in the special case where $F = U[0,2]$ and there are only two periods to sell. We walk through the key arguments in this special case; the formal details behind the claims made as part of this sketch appear as part of the proof of Theorem~\ref{thm:baseline} in the Appendix.

Given any discount factor $\delta$, Theorem~\ref{thm:baseline} yields the seller’s equilibrium prices:
\[
p_{1}^{*} = \frac{(2 - \delta)^{2}}{8 - 6\delta}, 
\quad 
p_{2}^{*} = \frac{2 - \delta}{8 - 6\delta}.
\]
In each period $t \in \{1,2\}$, the buyer learns whether $v > y_{t}^{*}$, where:
\[
y_{1}^{*} = \frac{4 - 2\delta}{4 - 3\delta}, 
\quad 
y_{2}^{*} = \frac{4 - 2\delta}{8 - 6\delta}.
\]
Buyers with $v > y_{1}^{*}$ purchase in period 1. Those with $y_{1}^{*} \ge v > y_{2}^{*}$ are indifferent between purchasing in period~1 or waiting until period 2, and in equilibrium purchase in period 2 (with ties broken against the seller). Buyers with $v \le y_{2}^{*}$ are indifferent between purchasing in period 2 or never, and in equilibrium never purchase. Since $F(y) = \frac{y}{2}$ for the uniform distribution, the seller's equilibrium profit is:
\[
\pi 
= p_{1}^{*} \left( 1 - F(y_{1}^{*}) \right) 
  + \delta p_{2}^{*} \left( F(y_{1}^{*}) - F(y_{2}^{*}) \right) 
= \boxed{\frac{(2-\delta)^{2}}{4(4-3\delta)}}.
\]
We now outline the key steps leading to this equilibrium.

\paragraph{\textbf{Step One: Worst-Case Information is Threshold Information in the Second Period.}}  
For any second-period price $p_{2}$, worst-case information solves an information design problem in the spirit of \citet{KG2011}: Nature chooses information to persuade the buyer not to buy. If $p_{2} \ge \mathbb{E}_{F^1_N}[v]$, the buyer would not purchase even without any additional information, yielding zero profit to the seller. We therefore focus on the case $p_{2} < \mathbb{E}_{F^1_N}[v]$ and $p_{2}$ lies above the lower bound of $\mathrm{supp}(F^1_N)$ (as otherwise, no possible information structure would deter purchase).

Since the buyer has a binary action set and the state space is continuous, \citet{Kolotilin_2015} implies that the solution involves revealing whether $v > y_{2}$, where $y_{2}$ satisfies\footnote{The case in which the posterior distribution has atoms is measure zero and is ignored here.}
\begin{equation} 
p_{2} 
= \int_{\underline{v}}^{y_{2}} v \, f(v \mid v \le y_{2}) \, dv.
\label{eq:BP}
\end{equation}
Under this threshold structure, the seller's more-preferred action (purchase) occurs when $v > y_{2}$, while the less-preferred action (no purchase) occurs when $v \le y_{2}$. The threshold is uniquely pinned down by the indifference condition: the buyer’s expected valuation conditional on a no-purchase recommendation must equal $p_{2}$. Any further attempt to reduce the purchase probability would require raising the buyer’s conditional expectation given a no-purchase signal, inducing purchase and making the deviation infeasible.\footnote{Relative to standard information design, an additional technical issue is that our analysis requires assumptions about off-path buyer behavior. Nature can provide strict incentives to the buyer while increasing the seller’s profit by only an arbitrarily small amount. The limiting profit must also be achieved in equilibrium; otherwise Nature would profitably deviate. Hence, the buyer breaks ties against the seller, even off path.}

\paragraph{\textbf{Step Two: Determining the Second-Period Price.}}  
Step One characterizes worst-case information for any $p_{2}$. Suppose the first period information structure also takes the form of a threshold: the buyer purchases immediately if $v$ exceeds this threshold, delaying otherwise.

Under this assumption, condition \eqref{eq:BP} and the fact that $F = U[0,2]$ imply that the profit-minimizing second-period threshold is $y_{2} = 2p_{2}$. Let $y_{1}(p_{1})$ denote the first-period sequentially worst-case threshold. Because the buyer purchases whenever $v > 2p_{2}$, the seller chooses $p_{2}$ to maximize expected profit
\[
p_{2} \left( 1 - \frac{2p_{2}}{y_{1}(p_{1})} \right),
\]
yielding the optimal second-period price
\[
p_{2}(p_{1}) = \frac{y_{1}(p_{1})}{4}.
\]

\paragraph{\textbf{Step Three: Determining the First-Period Indifference Point for the Buyer.}}  
In equilibrium, the function $p_{2}(p_{1})$ specifies the second-period price for \emph{any} on- or off-path $p_{1}$. By Step One, a buyer who delays is indifferent in period 2 between purchasing and not purchasing. Consequently, her \emph{expected} payoff at $t=1$ would be unchanged even if she were to \emph{always} purchase in the final period.

It follows that a buyer indifferent between immediate purchase and delay at $t=1$ is also indifferent between purchasing in period~1 and purchasing in period~2. Given a signal $s_{1}$, this indifference requires
\[
\mathbb{E}[v \mid s_{1}] - p_{1} 
= \delta \big( \mathbb{E}[v \mid s_{1}] - p_{2}(p_{1}) \big) 
\quad \Longrightarrow \quad 
\mathbb{E}[v \mid s_{1}] = \underbrace{\frac{p_{1} - \delta p_{2}(p_{1})}{1 - \delta}}_{=:w_{1}(p_{1})}.
\]
This characterization holds \emph{even off path}: in Step One, the solved worst-case information structure did not depend on whether the realized $p_{2}$ was chosen on path, and the same reasoning applies were the seller to deviate from the conjectured $p_{1}$.

\paragraph{\textbf{Step Four (Key Step): Finding the First-Period Information Structure.}}  
The problem of finding worst-case information in period 1 is now analogous to the period-2 problem in Step One. Seller profit is minimized if the buyer learns whether $v > 2w_{1}(p_{1})$. The first-period threshold is:
\[
y_{1}(p_{1}) = 2w_{1}(p_{1}).
\]
The key observation is that the seller’s payoff cannot be lowered beyond what this threshold strategy achieves: her optimal period-2 choice already ensures that a buyer delaying to period~2 is indifferent at the price $p_{2}(p_{1})$. Therefore, the buyer is indifferent between purchasing in period 1 and delaying \emph{if and only if} her expected valuation equals $w_{1}(p_{1})$. Even following a deviation of Nature, this conclusion remains given their last-period choice. 

Thus, the first-period worst-case information problem is essentially identical to that in Step One, except that the relevant indifference point is now $w_{1}(p_{1})$ instead of $p_{2}$.

\paragraph{\textbf{Step Five: Putting Everything Together to Find the Optimal First-Period Price.}}  
From Step~Two, $p_{2}(p_{1}) = y_{1}(p_{1})/4$, and from Step~Four, $y_{1}(p_{1}) = 2w_{1}(p_{1})$. Combining these yields
\[
p_{2}(p_{1}) = \frac{w_{1}(p_{1})}{2}.
\]
Substituting into the definition of $w_{1}(p_{1})$ from Step~Three gives
\[
w_{1}(p_{1}) = \frac{2p_{1}}{2 - \delta}.
\]
Given $p_{1}$, the seller’s expected profit is
\[
p_{1} \left( 1 - \frac{y_{1}(p_{1})}{2} \right) 
+ \delta p_{2}(p_{1}) \left( \frac{y_{1}(p_{1})}{2} - \frac{y_{2}(p_{1}, p_{2}(p_{1}))}{2} \right).
\]
Since $y_{1}(p_{1}) = 2 w_{1}(p_{1})$ and $y_{2}(p_{1}, p_{2}(p_{1})) = 2 p_{2}(p_{1}) = w_{1}(p_{1})$, this simplifies to
\[
p_{1} \left( 1 - \frac{2p_{1}}{2 - \delta} \right) 
+ \delta \frac{p_{1}}{2 - \delta} \left( \frac{2p_{1}}{2 - \delta} - \frac{p_{1}}{2 - \delta} \right).
\]
Maximizing over $p_{1}$ yields
\[
p_{1}^{*} = \frac{(2 - \delta)^{2}}{8 - 6\delta},
\]
as stated earlier. The equilibrium values $y_{1}^{*}$, $y_{2}^{*}$, and $p_{2}^{*}$ then follow directly.

\subsubsection{Discussion of the Solution} \label{sect:lessonone}\label{sect:solutiondiscussion}

One of the main economic takeaways from the above analysis is that the sequentially worst-case objective enables the seller to rely on simple pricing heuristics, avoiding much of the complexity involved in optimizing against \emph{arbitrary} learning processes.

The key observation---highlighted in Step Four---is that the value at which the buyer is indifferent between immediate purchase and delay depends \emph{only} on the conjectured price path, and not on the possibility of future learning. Consequently, the prospect of additional information in the second period does not increase the probability of delay, thereby justifying the first-period's threshold information structure  as worst-case. 

By contrast, if the buyer were to learn $v$ perfectly in period 2, say for exogenous reasons, the first-period worst-case threshold would generally need to be higher. In that case, any buyer with $\mathbb{E}_{F^1_N}[v] \le w_{1}(p_{1})$ would \emph{strictly} prefer to delay: knowing $v$ exactly in period~2 allows her to avoid purchasing when $v < p_{2}$, delivering additional surplus. The seller’s payoff could decrease if the first-period threshold were increased to induce more delay.

Our analysis therefore implies that sellers concerned about buyer learning can adopt strategies closely resembling those from the no-learning benchmark. In the general case with myopic threshold learning, the seller’s discounted expected profit from time $t$ onward (taking $w_{0} = y^{*}(w_{0}) = \overline{v}$) is:
\begin{equation*}
\sum_{s=t}^{T} \delta^{\,s-t} p_{s} \,
\frac{F\!\left(y^{*}(w_{s-1})\right) - F\!\left(y^{*}(w_{s})\right)}{F\!\left(y^{*}(w_{t-1})\right)}.
\end{equation*}
By comparison, when the buyer knows $v$ perfectly, the seller's discounted profit resembles this expression, with the only difference being that the purchase threshold in period each period $s \in \{t, \ldots, T\}$ is $w_{s}$ rather than $y^{*}(w_{s})$. Thus, differences in the seller's objective relative to the known-value case arise entirely through the function $y^{*}(\cdot)$, which can be computed to primitives.

\subsubsection{Additional Challenges in the Proof} 

The two-period case captures the basic intuition underlying Theorem \ref{thm:baseline}. The same logic applies generally: for sequentially worst-case information, the buyer cannot expect to obtain payoff-relevant information if delaying, thus implying the solution is a myopic threshold learning process. 

However, extending this reasoning to the general case is substantially more involved. Indeed, our exercise \emph{requires} solving for equilibrium strategies in a \emph{three-player} game (seller, buyer, Nature) for which establishing general existence and uniqueness is non-trivial.

We mention two technical issues in our proof beyond those discussed for the $T=2$ case:

\medskip
\noindent
\textbf{(1) Handling arbitrary past information structures.}  
In the informal construction, we solved for the second-period information structure \emph{assuming} the first-period information structure were partitional. This property need not hold in equilibrium (as other information
structures can induce buyer indifference following a recommendation to not buy). More significant, however, is that our general proof requires us to consider \emph{arbitrary} (past) information structures, which cannot be ruled out a priori---that is, before we determine what the worst-case actually is. Showing that the intuition from Section \ref{sect:lessonone} continues to hold in these general cases requires additional arguments. 

\medskip
\noindent
\textbf{(2) Allowing for on-path seller randomization.}  Our use of backward induction reasoning prevents us from imposing assumptions on the buyer's \emph{posterior} value distribution at any given time---in particular, we cannot make assumptions that would rule out ``early'' randomizations. Indeed, the posterior value distribution depends endogenously on the equilibrium learning process. Our proof accommodates the possibility that the seller might have randomized early, in anticipation of an exotic information structure arising later. While our theorem ultimately implies that on the equilibrium path seller randomization can occur only in the first period and is pinned down after that---consistent with the standard results in \citet{FLT1985} and \citet{GSW}---ruling out randomization in later periods is a substantive part of the argument.

\subsection{The Gap Case with $T=\infty$}

We now turn to the infinite-horizon case. Although the main focus of the paper is on the formulation of the problem and the structure of the sequentially worst-case learning process, it is natural as a theoretical exercise to compare our results with the standard Coasian bargaining benchmark. The main complication is that backward induction no longer applies in a straightforward way, since the standard skimming property is absent from our model. Nevertheless, when $\underline{v} > 0$, the traditional Coasian intuitions can still be recovered.

Readers familiar with the bargaining literature may associate the assumption $\underline{v} > 0$ with the conclusion that the market clears in finite time—i.e., that there exists a uniform bound $\hat{T}(\delta)$ such that, after any history, the buyer purchases with probability one by period $\hat{T}(\delta)$. This conclusion, however, requires additional regularity assumptions on the distribution of willingness-to-pay, such as Lipschitz continuity near the lower bound of its support (see \citet{GSW}). In our setting, such regularity conditions may fail for posterior distributions that arise under some learning processes, as we do not rule out any information structure a priori. 

It is straightforward to verify that the myopic threshold learning process can still be sustained as \emph{an} equilibrium strategy, by the one-shot deviation principle. The more subtle question is whether it is the \emph{unique} sequentially worst-case learning process among all equilibria. We show that uniqueness can be recovered under an intuitive restriction on the buyer’s equilibrium strategy.

\begin{definition}
An equilibrium $((\sigma,\alpha,\Pi),\mu)$ is a \textbf{monotone equilibrium} if and only if the buyer’s strategy $\alpha$ is weak-Markov—depending only on the posterior belief $F_B^t$ and the price $p_t$—and whenever $F_1$ strictly FOSD $F_2$, the condition $0 \in \alpha_t(F_1,p_t)$ implies $\alpha_t(F_2,p_t) = 0$.
\end{definition}

\noindent Monotonicity requires that if the buyer’s perceived distribution of willingness-to-pay becomes \emph{uniformly less} favorable, her willingness to delay cannot increase. It is immediate that when $T = \infty$, the equilibrium with the myopic threshold learning process is monotone. Moreover, under the same regularity assumption as in \citet{GSW}, we recover uniqueness.

\begin{proposition}  \label{prop:infhorizon}
Suppose $T = \infty$, $\underline{v} > 0$, and $F^{-1}$ is Lipschitz-continuous at $0$. Then the equilibrium price path is unique and deterministic (following possible randomization in period $t=1$). In any such equilibrium, every sequentially worst-case learning process induces the same buyer behavior as the myopic threshold learning process.
\end{proposition}

\noindent The proof shows that monotonicity restores the property that the market clears in finite time. Once this property holds, sequentially worst-case learning implies that future learning cannot be used to induce delay. Consequently, the seller’s value function at time $t$ can be written as:\footnote{This representation is guaranteed only on the equilibrium path under sequentially worst-case learning; it is therefore a \emph{result}, not an assumption, that it characterizes price-setting behavior. The proofs of Proposition~\ref{prop:infhorizon} and Theorem~\ref{thm:baseline} do not restrict learning processes and hence do not assume such a value function \emph{a priori}.}
\begin{equation}
V\big(y_{t-1}(p_{t-1})\big)
= \max_{p_{t}} \;
\Big[ p_{t} \big( F(y_{t-1}(p_{t-1})) - F(y_{t}(p_{t})) \big)
+ \delta \, V\big(y_{t}(p_{t})\big) \Big],
\end{equation}
with $y_{0} = \overline{v}$ and $y_{t}(\cdot)$ defined by~\eqref{eq:fullymyopicthreshold}.

Several structural properties of the equilibrium—such as the weak-Markov property\footnote{We call an equilibrium profile \emph{weak-Markov} if the buyer’s acceptance decision depends only on $F^t_B$ and $p_{t}$. This coincides with the standard definition when the buyer knows $v$.} and the absence of on-path seller randomization after the initial period—follow directly from the existence of this representation, by applying standard arguments from, for example, \citet{FLT1985} and \citet{AusubelDeneckere1989}.

\section{Beyond Sequentially Worst-Case via Safe Solutions} \label{sect:safe}

Theorem~\ref{thm:baseline} provides a sharp characterization of the sequentially worst-case learning process: it consists of descending partitional thresholds that make the buyer indifferent between purchasing and not purchasing in the absence of any further information. We now ask whether a seller who prices optimally against sequentially worst-case learning can achieve the same performance when \emph{arbitrary} learning processes are allowed. 

Specifically, our interest here concerns the robustness of a \emph{given} price path: we ask whether allowing for richer buyer learning possibilities---without changing the seller's pricing strategy---could lower the seller's payoff. Formally, this yields the following criterion: 

\begin{definition} \label{def:safe}
An equilibrium pricing strategy is a \textbf{safe solution} if, at any on-path history, the seller’s equilibrium payoff equals the worst-case profit obtained when using the \emph{same} pricing strategy against an \emph{arbitrary} learning process and any sequentially rational buyer strategy under it.
\end{definition}

\noindent We view this condition as natural even beyond the scope of our problem, but we restrict attention to our model to maintain focus. If the seller’s equilibrium pricing strategy is safe, then at any continuation history, the profit-guarantee against sequentially worst-case information coincides with the profit guarantee against \emph{arbitrary} learning. In particular, this applies in the initial period: richer learning possibilities would not lower seller profit below the level identified in Theorem \ref{thm:baseline} whenever the seller follows the equilibrium price path. Thus, the equilibrium solution retains the normative appeal that often motivates robust objectives. 

In other words, if a sequentially worst-case price-path is safe, then a seller who is self-aware of his lack of commitment, and anticipating that he will reason similarly in every future period, would maintain their same pricing strategy. Since we apply this criterion at any on-path history, we implicitly assume that the seller does not revisit ``worst-case'' scenarios from earlier periods.\footnote{An asymmetry arises when modifying past learning to hurt the seller, since any history is conditional on no purchase---implicitly assuming that signals were (relatively) unfavorable. By contrast, discouraging the buyer from buying following a price drop will may require some information to be provided that would encourage purchase, unable to condition on an unfavorable realization. We discuss implications of this possibility in Appendix \ref{sect:worsepast}.} It is also important to note that safe solutions \emph{allow} the seller to possibly be hurt under some \emph{joint} change in the price path and the learning process. Indeed, one can typically construct a learning process such that the pricing strategy induced as an equilibrium strategy holding fixed this (non–sequentially worst-case) learning process yields a lower payoff than the benchmark in Theorem \ref{thm:baseline}. Intuitively, this occurs when the learning process transfers additional surplus to the buyer to induce further delay. Example \ref{worseinfoexample} in Appendix \ref{sect:alternatives} illustrates how such transfers may result in a lower seller payoff. This observation underscores the subtleties inherent in the concept.

\medskip

  The following property is sufficient for the price path in Theorem~\ref{thm:baseline} to be safe.

\begin{definition}
Let $y(w)$ satisfy $w = \mathbb{E}_{v \sim F}[v \mid v \le y(w)]$. We say that a distribution $F$ is \emph{threshold-ratio monotone} if 
\[
\frac{y(w)}{w} \quad \text{is weakly increasing in } w.
\]
\end{definition}

\noindent Recall that the myopic threshold learning process in Theorem \ref{thm:baseline} informs the buyer whether $v > y(w_{t}(p_{t}))$. Intuitively, threshold-ratio monotonicity ensures that increasing the threshold to induce more delay does not disproportionately raise the conditional expectation $\mathbb{E}[v \mid v \le y]$. In other words, the expected value of buyers below the threshold in earlier periods changes less than the expected value in later periods. This guarantees that any reduction in early-period revenue is smaller than the corresponding gain from later-period sales, as our second main Theorem shows: 

\begin{theorem} \label{thm:selfconfirming}
Suppose the value distribution is threshold-ratio monotone. Then the equilibrium pricing strategy in Theorem~\ref{thm:baseline} and Proposition~\ref{prop:infhorizon} is a safe solution: if the seller uses the strategy described there, no learning process can reduce the seller’s expected payoff at any on-path history.
\end{theorem}

%Note that, on the equilibrium path, the only event the seller observes is that the buyer has not yet purchased the product. We assume the seller does not revisit “worst-case” scenarios from earlier periods, but rather takes past events as given. Thus, the expected payoff is only ``safe'' for any \emph{on-path} history.

Theorem~\ref{thm:selfconfirming} explicitly solves for the worst-case learning process over all possible processes under the assumption of threshold-ratio monotonicity, and shows that this process coincides with the one in Theorem~\ref{thm:baseline}. The key difference from the exercise of finding the sequentially worst-case process lies in the conjectures the \emph{buyer} may form regarding the learning process. Sequentially worst-case constraints matter because the seller recognizes that the buyer will not anticipate a learning process that is not itself sequentially worst-case. In Theorem~\ref{thm:selfconfirming}, by contrast, we fix the price path and, in principle, allow more information to be provided to the buyer to induce additional delay. The proof relies on the fact that the equilibrium price path is deterministic (after the initial period), but otherwise places no restriction on the seller’s strategy.

A useful preliminary observation is that, given a fixed price path, the worst-case learning process is partitional in each period. Even so, substantial work remains because the worst-case thresholds may differ from those implied by the sequentially worst-case process. Recall that sequentially worst-case requires that whenever the buyer delays, she is indifferent between delay and purchase. Put differently, threshold information structures could still make the buyer \emph{strictly} prefer to delay. Thus, solving for worst-case information involves a non-trivial choice of a threshold for each period, subject to the buyer’s obedience constraints.

We address this by identifying a specific adjustment of the partition thresholds that lowers discounted profit whenever a threshold fails to induce exact indifference after a recommendation not to buy. While lowering the threshold increases sales in that period, we adjust the previous period's threshold to preserve obedience. In the Appendix, we verify that, under threshold-ratio monotonicity, this adjustment strictly reduces the seller’s profit.

While not every distribution satisfies threshold-ratio monotonicity, the condition is still quite permissive. The following sufficient condition illustrates this point.

\begin{proposition} \label{prop:suffcond} 
For differentiable $f$, threshold-ratio monotonicity holds if $\frac{v f(v)}{F(v)}$ is decreasing in $v$.
\end{proposition}

\noindent  Proposition \ref{prop:suffcond} implies that the class of threshold-ratio monotone distributions includes $F(v)=v^{a}$ for $a >0$, exactly the case considered in \cite{FS2013}, for instance. Threshold-ratio monotonicity also holds for uniform distributions supported on compact intervals of $\mathbb{R}^+$. Consequently, the price path described in Section \ref{sect:twoperiods} is safe.

Finally, if threshold-ratio monotonicity holds, the restriction to monotone equilibria in Proposition~\ref{prop:infhorizon} can be substantially relaxed.

\begin{definition}
An equilibrium $((\sigma,\alpha,\Pi),\mu)$ is a \textbf{deterministic equilibrium} if and only if the equilibrium pricing strategy is deterministic except possibly in the first period.
\end{definition}

\noindent The actual price path from the second period onward may depend on the realization of the first-period price. Equilibria in the known-values case under the Lipschitz condition satisfy this restriction (see \citet{FLT1985} and \citet{GSW}).\footnote{\citet{GSW} also discuss why the equilibrium pricing strategy may not be deterministic in the initial period, hence our inclusion of this condition in the definition.}

\begin{corollary} \label{cor:unidete}
If, in addition to the conditions of Proposition \ref{prop:infhorizon}, $F$ is threshold-ratio monotone, then the equilibrium in Proposition \ref{prop:infhorizon} is the unique deterministic equilibrium.
\end{corollary}

\section{Saddle-Point Worst-Case Formulations} \label{sect:saddle}

We now discuss an equivalent formulation of our problem. Our reason for doing so stems from our use of ``Nature as a player'' as an expositional device. Analytically, this step is useful as it motivates our solution concept, which can be analyzed using well-known methods. But this formulation is also conceptually useful since once equilibrium is imposed, (i) information minimizes seller profit on- and off-path, and (ii) the seller correctly anticipates these choices so that dynamic consistency holds. We view the economic content of our model to be clearer when described in the extensive form, analogous to classic works like \cite{FLT1985} and \cite{GSW}. Nevertheless, it is also instructive to formulate a normal-form version of this interaction to shed further light on its interpretation and connection to past work. 

While this expositional device is often invoked in static settings, the existence of a saddle-point solution---where the designer's choice, say $x \in X$, facing uncertainty over $y \in Y$, solves $\max_{x \in X} \min_{y \in Y} u(x,y)$---implies that ``Nature'' need not be interpreted literally. It simply turns out that the optimal choices under this objective coincide with the equilibrium of a game where Nature chooses $y \in Y$. We show the same conclusion holds in our framework by identifying the analogous saddle-point formulation. Thus, the difference with static worst-case formulations is \emph{not} that our appeal to Nature as a player is more literal, but rather in terms of the possibilities considered. These observations clarify more precisely the role of Nature in our formulation.

\subsection{Sequentially Worst-Case Learning} \label{sect:mainnormal}

To motivate the formulation, consider the no-learning version of our problem; say, where the buyer knows $v$ at time 0. In this case, the equilibrium outcome delivers a pricing strategy, say $(p_{1}^{*}, p_{2}^{*}, \ldots)$. This same price path emerges in \emph{Nash} equilibrium if, instead of choosing a new price at every point in time, the seller chose a price path \emph{at time 0} with future prices ``hidden''---so that the buyer only observes the price in a given period once that period arrives. In this case, the seller chooses $p_{t}$ conditional on the buyer not having bought by time $t$. Thus, consider the following normal form game, which we refer to as the \emph{hidden-price-hidden-information normal form}: 

\begin{itemize} 
\item The seller's action set is the set of all randomized pricing strategy $\sigma : \cup_{t} H_{S}^{t} \rightarrow  \Delta(\R_{+})$.

\item Nature's action set is the set of all history-dependent learning processes
\[
\Pi : \bigcup_{t} H_{N}^{t} \to \Delta\left(\left\{(\pi,S)\right\}\right),
\]
\item After the seller and Nature simultaneously choose their actions, the buyer chooses $\alpha$, given $\mu$ (with $\mu$ derived from Bayes rule given $\sigma$ and $\Pi$), for all $h_{B}^{t}$, to maximize the continuation payoff conditional on reaching $h_{B}^{t}$:
\[
\mathbb{E}_{\mu,\mathbb{P}_{\sigma,\alpha,\Pi}}\!\left[\sum_{\tau \ge t} \delta^{\tau-t} (v-p_{\tau})\,\mathbf{1}_{\{\text{accept at }\tau\}} \,\big|\, h_{B}^{t}\right],
\]
where $\tau$ is the induced stopping time.
\end{itemize}

\noindent Crucially, the buyer's strategy here depends only on $h_{B}^{t}$, the set of prices and information observed \emph{up until time } $t$. It \emph{does not} include future prices or future information---buyer actions at time $t$ cannot condition on choices of other players at time $t+s$. We have the following: 

\begin{lemma}
\label{lem:1}
The Nash equilibrium in the hidden-price-hidden information normal form induces the same equilibrium outcome as in the equilibrium of the main model.
\end{lemma}

Lemma \ref{lem:1} provides a mapping between our worst-case formulation and those from static robustness settings, where optimal outcomes correspond to a saddle point: choices are optimal against the worst-case realizations from a given set of possibilities. The same conclusion holds in the hidden price-path formulation of the model, simply with the additional requirement that future information structures cannot be revealed to the buyer to shape her strategy. While described as a Nash equilibrium outcome, as mentioned above, the implication is that the solution corresponds to a saddle-point of an objective for which no such literal interpretation is necessary. Ultimately, the sequentially worst-case benchmark requires only that future information is ``hidden.''

\subsection{Safe Solutions} 
We now show that safe solutions emerge naturally in the saddle-point formulation when the restrictions in the previous paragraph are dropped: 

\begin{definition}
\label{def:one-shot}
A profile $(\sigma, \alpha, \Pi)$ is \textbf{one-shot worst-case} learning process $\Pi$ minimizes 
\begin{equation}
\mathbb{E}_{\mu,\mathbb{P}_{\sigma,\alpha,\Pi}}\!\left[\sum_{k=t}^{T} \delta^{k-t} p_{k}\,\mathbf{1}_{\{a_{k}=1\}} \ \right],
\end{equation}
i.e., the ex-ante expected seller's discounted payoff, and the buyer's strategy is optimal at every $h_{B}^{t}$ given the realized $\Pi$. 
\end{definition}
While this notion maintains the assumption from the previous section that the price is revealed to the buyer period-by-period, the key difference is that it allows the information arrival process to be revealed to the buyer \emph{at the very beginning}. Thus, given $\Pi$, the game becomes a two-player game between the seller and the buyer. 

Motivated by this observation, we consider the normal-form game that arises under this modification, which we call the \emph{hidden-price-observed-information normal form}: 

\begin{itemize} 
\item The seller's action set is the set of all randomized pricing strategy $\sigma : \cup_{t} H_{S}^{t} \rightarrow  \Delta(\R_{+})$.

\item Nature's action set is the set of all history-dependent learning processes
\[
\Pi : \bigcup_{t} H_{N}^{t} \to \Delta\left(\left\{(\pi,S)\right\}\right),
\]

\item The buyer chooses $\alpha$, given $\mu$ \textbf{and $\Pi$}, for all $h_{B}^{t}$, maximizing the continuation payoff conditional on reaching $h_{B}^{t}$:
\[
\mathbb{E}_{\mu,\mathbb{P}_{\sigma,\alpha,\Pi}}\!\left[\sum_{\tau \ge t} \delta^{\tau-t} (v-p_{\tau})\,\mathbf{1}_{\{\text{accept at }\tau\}} \,\big|\, h_{B}^{t}\right],
\]
where $\tau$ is the induced stopping time.
\end{itemize}

The difference with the hidden-information case in Section \ref{sect:mainnormal} arises since when $\Pi$ is chosen, this choice is observed by the buyer as suggested by the name of this normal form. Note, however, that we still assume simultaneity in the choice of pricing strategy and the choice of learning process. We have the following lemma:
\begin{lemma}
\label{lem:2}
The Nash equilibrium in the hidden-price-observed-information normal form game (between the seller and Nature) induces a one-shot worst-case learning process.
A sequentially worst-case equilibrium pricing strategy is a safe solution only if it can be sustained in some Nash equilibrium of the hidden-price-observed-information normal form game.
\end{lemma}
Taken together, these results shed new light on both the notions of sequentially worst-case learning process and safe solutions---and their connection as articulated in Theorem \ref{thm:selfconfirming}. In particular, Lemma \ref{lem:1} shows that the sequentially worst-case benchmark arises from restricting the set of possibilities over which the seller considers worst-case outcomes, rather than representing a literal opponent per se, while Lemma \ref{lem:2} shows that safe solutions arise from removing these restrictions. As such saddle-point formulations have come up naturally in the analysis of static contexts, our hope is that articulating these benchmarks connects our assumptions to those in past work.

\section{Conclusion} 

\subsection{Outcomes under Equilibrium with Fixed Learning Process} \label{sect:positiveselection}

Part of the motivation for our robust approach is the observation that classical Bayesian approaches have tended to only analyze particular forms of learning processes. Here, we argue that insofar as this could be allowed, the benchmark results would likely resemble ``anything goes'' more than anything sharp. Specifically, allowing for general unrestricted leaning processes without imposing some restriction (such as our sequentially worst-case) may yield dramatic departures from the known-values equilibrium.\footnote{This exercise is in the spirit of \emph{robust predictions}; \cite{Liu2022} performs such an exercise when the \emph{seller} may obtain extra information about the buyer's value, showing that a rich set of payoffs may emerge in the frequent-offer limit.} The following shows that indeed this possibility enables constant-price path equilibria to emerge for some fixed learning process:

%\textcolor{red}{I don't think this proposition 2 is helpful. First in this equilibrium the seller does better compared to fully myopic, it doesn't make too much sense here. Second we already have proposition 3. Thirdly, the only thing special about this proposition is the constant path price, I know you want to compare with commitment case? But the thing is constant path will remind people of coarse conjecture, not good. I think if you want you can put it in the appendix, not main text. }

\begin{proposition} \label{prop:unboundedTime}
Fix $F$ and $\delta$, and take $T = \infty$. Suppose the equilibrium outcome when the buyer knows $v$ does not involve market clearing at $t = 1$. Then there exists a fixed learning process admitting an equilibrium (between the seller and the buyer under this process) such that:  

\begin{itemize} 
\item The seller follows a constant price path.  

\item The seller's expected payoff is $v^{*}$, where $v^{*}$ is any value strictly less than $\E_{F}[v]$ and strictly greater than the equilibrium payoff identified in Proposition \ref{prop:infhorizon}.  

\item The market does not clear in any finite time (i.e., there is no $\hat{T}$ such that the buyer purchases before time $\hat{T}$ with probability one on-path).  
\end{itemize}

\noindent Moreover, when $T < \infty$, there exists a fixed learning process and an equilibrium satisfying the first two points as long as $v^{*}$ is strictly greater than the equilibrium payoff identified in Theorem \ref{thm:baseline}.

\end{proposition}

\noindent This result highlights (i) the possibility of equilibrium \emph{multiplicity} for a \emph{fixed} learning process, and (ii) the absence of a finite time horizon by which the market clears, neither of which arises in the known-values gap case.\footnote{The combination with a constant price path is also distinct, although constant prices would arise in the degenerate case where the market clears at time 1.} Proposition \ref{prop:unboundedTime} further shows that if arbitrary learning process is possible, then severe departures from the known-values predictions can emerge.

The learning process used to prove Proposition \ref{prop:unboundedTime} is surprisingly simple: On path, the seller sets price equal to $\mathbb{E}_{F}[v]$, no information is provided if the seller adheres to the constant price path, and a deviation by the seller causes the buyer to learn $v$ perfectly. Since the buyer's expected payoff is 0 at every time, this construction supports an equilibrium in which the buyer randomizes with a probability that induces the seller to maintain the constant price path rather than deviating. In particular, triggering the release of information can lead to highly unfavorable outcomes for the seller. The key property of this learning process is that the prospect of learning \emph{does} shape equilibrium outcomes, as it prevents the seller from deviating.

The takeaway from this discussion is that our robust approach has an appealing feature: it recovers much of the usual intuition from the well-studied literature on known-values bargaining. It is worth emphasizing that a conclusion as sharp as the one we derive need not hold under alternative formulations of the robust objective, a point discussed in detail in Appendix \ref{sect:alternatives}.\footnote{For example, naivete may lead the seller not to sell at all if she expects the buyer to have strong delay incentives.}

\subsection{Discussion of the No-Gap Case} \label{sect:nogap}

While our main analysis focused on either the finite-horizon or gap cases, many of our main insights also apply to the no-gap case:

\begin{proposition}\label{prop:2}
Suppose $T = \infty$ and $\underline{v} = 0$. An equilibrium exists in which Nature implements myopic threshold learning process.
\end{proposition}

Given that it is typically impossible to impose finite-time market clearing when $\underline{v} = 0$, monotonicity alone does not suffice to rule out non-myopic-threshold learning processes. However, the threshold-ratio monotonicity condition on $F$ does imply that this outcome belongs to the same class of equilibria, with possible multiplicity:

\begin{proposition}\label{prop:nogapunobs}
Suppose $T = \infty$ and $\underline{v} = 0$. If $F$ is threshold-ratio monotone, then in any deterministic equilibrium Nature implements myopic threshold learning process.
\end{proposition}

\noindent This proposition highlights the relevance of myopic-threshold learning process even in the no-gap case. Note that in the no-gap case we no longer obtain unique equilibrium price paths, just as in the known-values case—identical constructions such as in \cite{AusubelDeneckere1989} apply once Nature's strategy is restricted to being myopic threshold learning process.

\subsection{Discussion of Model Assumptions} \label{subsect:discussion}
We defer a detailed exploration of alternative formulations to Appendix \ref{sect:alternatives}. The explicit use of ``Nature'' as a player is primarily an expositional device, clarifying why one might expect dynamic consistency of the learning process to be preserved. In equilibrium, actions must maximize payoffs given that future actions are determined by the equilibrium profile, which in turn must satisfy the same requirement. While ours is an incomplete-information game (complicating backward induction), this reasoning provides the key intuition. In particular, equilibrium requires that the learning process the seller anticipates is precisely the one that materializes both on-path and after deviations—there is no ambiguity regarding which learning process the seller should consider at time $10$. Section \ref{sect:saddle} further underscores that the sequentially worst-case learning process also emerges under a traditional (static) formulation of the worst case.

Our introduction contrasted our use of a maxmin objective with earlier approaches in the literature. \cite{ES_2003} propose a ``rectangularity'' condition on the set of priors that characterizes when the maxmin decision rule is dynamically consistent. While \cite{ES_2003} also provide a procedure for constructing a set of priors to represent a decision maker's ex-ante preference, their setting differs from ours in part due to the presence of the buyer. That said, we acknowledge that our exercise is in spirit similar to their proposal. We simply find it more direct, in our setting, to relax assumptions on ``Nature's commitment power'' rather than on the set of learning processes and (endogenous) buyer equilibrium strategies.\footnote{Another approach to accommodating ambiguity aversion is introduced in \cite{klibanoff2009recursive}, which develops a dynamically consistent version of smooth ambiguity preferences allowing for a richer separation between ambiguity and ambiguity attitudes than maxmin. \cite{hanany2020incomplete} extend this framework to games.}

On this note, we have in mind situations where the buyer can consult any source that comes to mind costlessly. The restrictions on what the buyer knows about $v$ reflect limits on what the buyer has access to at any time, rather than a choice to exert less effort. We share this assumption about how information is generated with much of the informational robustness literature (e.g., \cite{Du2018,FirstBrooksDu,BrooksDuDuality,DR2023}, among others). 

A general issue for robust objectives concerns the timing of the worst case. We have in mind situations where buyers have some time to respond to offers, or where new offers are made very shortly after rejection (so that the buyer can learn while considering the offer). While the seller may randomize, the information realized in a given period can depend on the seller's actions (i.e., the posted price), insofar as such actions lead the buyer to obtain some other information source. Indeed, many papers have noted channels through which information can depend on price in practice—see, for instance, \cite{XuYang_2022,LBC2023,IS_2023}. \cite{KeZhang2020} provide decision-theoretic foundations for the assumption that the worst case depends on realized choices, an assumption also adopted in other work applying such approaches (e.g., \cite{Carroll_2015,BBM2017,Chen2023,GS2023,Malladi2023}).

For us, allowing the worst case to condition on realized seller choices is more compelling for three reasons. First, allowing information to depend on the randomization but not on its realization evokes commitment, since it requires a seller who observes the outcome of the randomization not to reconsider (for instance, after observing the price draw but before posting it). Our interest, however, is in cases where no commitment ability is present. The inability to commit to a randomization has traditionally been used to justify restricting attention to deterministic mechanisms more generally \cite[see][p.~67]{LaffontMartimort}. Second, assuming that information is independent of realized seller choices implies that information cannot react to \emph{deviations}. Yet a seller may well be deterred from deviating by the prospect of directly influencing the informational environment (e.g., attracting attention through unexpected actions). Allowing some price dependence therefore seems natural. As we see no universally plausible a priori restriction, we therefore leave this possibility unconstrained. Third, given that our goal is to study a seller completely unsure of how their actions influence buyer learning, it seems reasonable to start with the case where Nature has as much power as possible.

\subsection{Final Remarks}

In this paper, we propose a new approach to modeling a seller who reoptimizes a dynamic worst-case objective. By treating adversarial Nature as a strategic player in a dynamic game, we obtain a dynamically consistent worst-case objective and sharp characterizations of equilibrium outcomes. A seller of a house, for instance, need not worry about delay caused by the buyer wishing to consult a family member: the sequentially worst-case learning process simply minimizes the seller’s payoff period by period, given conjectures about how prices will evolve. Even when other learning processes are possible, in many cases this is immaterial, as no prospective future information would further reduce the seller’s payoff.

Durable goods pricing is a natural first setting to study informational robustness without commitment, as the buyer’s problem reduces to choosing \emph{when} to purchase. Our analysis also uncovers conceptual issues that arise more broadly when relaxing commitment in a robust framework. To clarify these, Appendix \ref{sect:alternatives} considers alternative specifications of the robust objective, arguing that our formulation yields the most intuitive and tractable solution for informationally robust sales.\footnote{For example, a seller who considers the worst case over all learning processes but does not anticipate that the process may change over time might never attempt to sell, even if only moderately patient; see Appendix \ref{sect:naive}.} We do not claim this conclusion extends to all environments, but in our application the potential for dynamic inconsistency appears less severe than previously thought; when safe solutions exist, as they do for a large class of environments, allowing for potentially dynamically inconsistent learning processes would not hurt a dynamically-consistent seller. 

We view our contribution to the agenda outlined by \cite{BergemannValimakiSurvey} as presenting a tractable dynamically consistent informationally-robust objective. We also aim to offer a template for extending the robust approach to other dynamic interactions. While one might argue that focusing on a restricted worst case departs from the fully robust objective, we have shown that in our setting this objection often has little bite. By introducing the notion of a \emph{safe solution}, we aim to facilitate tractable solutions in other dynamic models and support the case that such solutions can remain consistent with the original motivation for adopting a robust approach.

\newpage

\bibliographystyle{ACM-Reference-Format}
\bibliography{coase}

\appendix

\newpage

\section{Proofs}

\subsection{Characterization of Sequentially Worst-Case}
\begin{proof}[Proof of Lemma \ref{lem:1}]
We prove the result by backward induction. Let the Nash equilibrium strategy be denoted by $(\sigma^*, \Pi^*)$. At time $t = T$, since Nature can secretly deviate, for any $p_T$ in the support of $\sigma^*(h_S^T)$ the information structure $\Pi^*(h_N^T)$ must minimize the seller's expected payoff. For off-path $p_T$, it is also without loss of generality to assume that $\Pi^*(h_N^T)$ minimizes the seller's expected payoff, as this does not affect the equilibrium outcome. Given Nature's strategy $\Pi^*$, since the seller can also secretly deviate at $t = T$, $\sigma^*(h_S^T)$ must maximize the seller's expected payoff given $\Pi^*$. Repeating this reasoning recursively, the conclusion follows by backward induction.
\end{proof}

\begin{proof}[Proof of Lemma \ref{lem:2}]
In the Nash equilibrium, the learning process must minimize the seller's ex-ante profit as otherwise Nature can secretly deviate to another learning process to hurt the seller even more as this deviation will be observed by the buyer. Now if a sequentially worst-case learning process is a safe solution, then by the definition Nature's learning process on the equilibrium path is a one-shot worst-case learning process. And because the seller and buyer are both sequentially rational on the equilibrium path, we conclude it is a Nash equilibrium between the seller and Nature. 
\end{proof}

\begin{proof}[Proof of Theorem \ref{thm:baseline}]\
\begin{enumerate}[label=\textbf{Step \arabic*.}, wide=0pt, leftmargin=*, align=left, labelwidth=*, itemsep=1ex, listparindent=\parindent, parsep=0pt]

\item \textbf{Reduction to Binary Recommendations.}  
We first show that, without loss of generality, Nature can be assumed to provide only \emph{binary} recommendations to the buyer in each period—``buy'' or ``wait.'' This is the familiar \emph{``recommendation principle''} from the information design literature.

\begin{lemma} \label{lemm:revprinciple}
Given equilibrium strategies $(\sigma, \alpha, \Pi)$ and the corresponding belief system $\mu$, there exist equilibrium strategies $(\tilde{\sigma}, \tilde{\alpha}, \tilde{\Pi})$ and a belief system $\tilde{\mu}$ such that $\tilde{\Pi}$ provides only two signals (\emph{buy} or \emph{wait}) at every history $h_N^t$, and the induced equilibrium outcome is identical to that under $(\sigma, \alpha, \Pi, \mu)$.
\end{lemma} 

\begin{proof}
Let $(\pi_t, S_t) = \Pi(h_N^t)$ denote the information structure chosen by Nature at history $h_N^t$.  
With a slight abuse of notation, write $\alpha(h_B^t, s_t)$ for the probability that the buyer purchases after observing $s_t \in S_t$ in period $t$.

\smallskip
We first show that it is without loss to assume the buyer doesn't randomize. If $\alpha(h_B^t, s_t) \in (0,1)$ for some $s_t$, Nature can ``refine'' the signal to implement the buyer’s mixed action deterministically.  
Specifically, construct a new information structure $(\tilde{\pi}_t, \tilde{S}_t)$ that, conditional on $s_t$, sends:
\[
\overline{s}_t \quad\text{with probability}\quad \alpha(h_B^t, s_t),
\quad\text{and}\quad
\underline{s}_t \quad\text{otherwise},
\]
where the buyer always buys after $\overline{s}_t$ and never buys after $\underline{s}_t$.  
Let $\tilde{\alpha}$ coincide with $\alpha$ except that it is now deterministic.  
This transformation preserves all outcomes and incentives.  
Applying this to each $(t,s_t)$, we may assume $\alpha(\cdot)$ is deterministic.

\smallskip
With $\alpha$ deterministic, all signals on which the buyer buys can be pooled into a single ``buy'' signal $\overline{s}_t$, and all others into a single ``wait'' signal $\underline{s}_t$.  
Formally, replace $\pi_t$ by
\[
\tilde{\pi}_t : [\underline{v},\overline{v}] \to \Delta(\{\underline{s}_t, \overline{s}_t\}),
\]
where $\overline{s}_t$ is sent if and only if the original $\alpha$ prescribes purchase. To retain any finer distinctions lost by pooling at period $t$ that might matter in later periods, Nature can carry forward the original detailed signals at period $t$ to period $t+1$ if needed. Formally:
\[
\tilde{\Pi}(h_N^t) = (\tilde{\pi}_t, \{\underline{s}_t, \overline{s}_t\}),
\]
and after $(h_N^t, \tilde{\pi}_t, \tilde{s}_t, p_{t+1})$, Nature delivers the continuation information $(\pi_t, S_t) \cup \Pi(h_N^t, \pi_t, s_t, p_{t+1})$.

\smallskip
This construction changes neither the buyer’s nor the seller’s behavior, and the induced outcome is identical to that under $(\sigma, \alpha, \Pi, \mu)$.  
By induction on $t$, we conclude that Nature can be restricted, without loss, to binary recommendations at every history.
\end{proof}

\medskip 

\item \textbf{Notation and Preliminaries.}  
Given Lemma~\ref{lemm:revprinciple}, we may assume without loss that Nature provides binary recommendations in every period. We now introduce notation used throughout the remainder of the proof.

\medskip
\noindent\textit{Histories and posterior bounds.}  
Fix a seller history $h_S^t$, and let $\underline{v}_S^t$ be the infimum of the support of $F_S^t$, the posterior distribution of $v$ conditional on $h_S^t$.  

\medskip
\noindent\textit{Unconditional CDFs.}  
For convenience, we express all Bellman iterations in terms of \emph{unconditional} probabilities. Let
\[
G_S^t(\cdot)
\]
denote the CDF associated with a general Radon measure on $[\underline{v},\overline{v}]$.  
Note that $G_S^t(\overline{v})$ need not equal $1$; rather, it can be scaled to $F_S^t(\cdot)$ divided by the remaining buyer mass.\footnote{Formally, we work in the space $\text{Radon}([\underline v, \overline v])$ with the weak-$*$ topology.}

\medskip
\noindent\textit{Threshold information structures.}  
Given a history $h_S^t$ and a scalar $x$, define
\begin{equation}
y_S^t(x) 
:= \inf \big\{ y \ge \underline{v}_S^t \;:\; x \le \mathbb{E}_{F_S^t}[\,v \mid v \le y\,] \big\}.
\label{eq:thresholdexpression}
\end{equation}
Intuitively, $y_S^t(x)$ is the threshold type below which the buyer’s expected valuation is at most $x$.

Since $F$ has a well-defined density, and—by Lemma~\ref{lemm:revprinciple}—we may restrict attention to binary recommendations, $F_S^t$ is continuous everywhere except possibly at histories where the buyer is certain to purchase. In particular, in \eqref{eq:thresholdexpression} the inequality either holds with equality or yields $y_S^t(x) = \underline{v}_S^t$. Moreover, $y_S^t(\cdot)$ is continuous in $x$.

\medskip
\noindent\textit{Buyer cutoff.}  
Let $w_t = w_t(h_B^t)$ be the buyer expected valuation indifferent between buying immediately at $t$ and delaying. This is purely for expositional convenience: at this point we have not established the skimming property or characterized the equilibrium. In principle, a low-posterior buyer could expect future information that increases her valuation.

\medskip
\noindent\textit{Seller best-response correspondences.}  
Define:
\[
M_t(G_S^t) = \arg\max_{p_t} V^t(G_S^t(\cdot))
\]
and
\[
m_S^t(b) = \arg\max_{p_{t+1}} V^{t+1}(G_S^{b,t}(\cdot)),
\]
where $G_S^{b,t}(\cdot)$ is the CDF obtained by truncating $G_S^t(\cdot)$ at $b$, and $V^t(\cdot)$ is the seller’s expected payoff in period $t$.\footnote{Since uniqueness is not yet established, $V^t(\cdot)$ need not be uniquely defined, and the weak-Markov property need not hold; the continuation value may depend on both the distribution and the full history. We therefore reference a particular equilibrium throughout.}

\medskip 
\item \textbf{Backward induction}  
We now proceed by backward induction. The general strategy is as follows:  
since, along the equilibrium path, the buyer’s ex-ante value of future information is zero, Nature’s best response in any given period is to \emph{minimize} the probability of purchase in that period, as the equilibrium price path is decreasing.

\subsubsection*{Nature’s last-period problem}
Consider $t = T$.

\smallskip
\noindent\textit{Case 1: Market already cleared.}  
If the market has already cleared, any strategy is optimal for Nature; in particular, she can provide a trivial “no information” structure.

\smallskip
\noindent\textit{Case 2: Market not cleared.}  
Let $F_S^T$ denote the posterior distribution of $v$ among remaining buyers, with unconditional CDF $G_S^T$.  

If $p_T \le \underline{v}_S^T$, the buyer must purchase with probability~1 in equilibrium, regardless of Nature’s choice.\footnote{This is a standard argument in the Coasian bargaining literature; see \citet{FLT1985}.}  
In this case, the claim that Nature uses a threshold information structure giving the same buyer payoff as no information is trivially satisfied.

If instead $p_T > \underline{v}_S^T$, Step One of Section~\ref{sect:twoperiods} implies the worst-case information structure is a \emph{threshold} with cutoff $y_S^T(p_T)$. Under such a structure, one of two things happens:
\begin{enumerate}
\item The buyer is indifferent between purchasing and not when recommended to wait, or
\item Nature’s choice has no effect on the buyer’s action (e.g., if $p_T$ is too high).
\end{enumerate}
In either case, the buyer’s expected payoff matches that under no new information. By sequential rationality, Nature’s equilibrium last-period strategy at any history $h_N^T$ is therefore a threshold rule. For any $p_T > \underline{v}_S^T$, Nature can approximate the worst-case profit while giving the buyer a \emph{strict} incentive to buy by, e.g., revealing whether $v \le y_S^T(p_T) - \varepsilon$ for arbitrarily small $\varepsilon > 0$.  
Therefore, in equilibrium, the buyer must break indifference \emph{against} the seller; otherwise Nature could profitably deviate.

\subsubsection*{Seller’s last-period problem}
Consider the history $p^{T-1}$.

\smallskip
\noindent\textit{Case 1: Market should have cleared but hasn’t.}  
If, under $(\Pi, \alpha)$ (with Nature and the buyer following equilibrium strategies, but allowing possible seller deviations), the market should have cleared yet continues, this history is off-path. We may assign arbitrary beliefs and prices to deter such deviations.\footnote{For example, the seller could post $\overline{v}$.}

\smallskip
\noindent\textit{Case 2: Market not cleared on path.}  
If the market should not have cleared and indeed has not, then the only equilibrium-consistent belief for the seller is that Nature and the buyer have followed $(\Pi, \alpha)$. Thus the remaining posterior is $F_S^T$ with unconditional CDF $G_S^T$.

For any $p_T$ in the support of $\sigma_T$ we must have:
\[
p_T \in M_T(G_S^T) 
= \arg\max_{p_T} V^T(G_S^T(\cdot)) 
= \arg\max_{p_T} p_T\left(G_S^T(\overline{v}) - G_S^T\big(y_S^T(p_T)\big)\right).
\]
Since both $G_S^T(\cdot)$ and $y_S^T(\cdot)$ are continuous, the objective is jointly continuous in $(G_S^T, p_T)$. By Berge’s Maximum Theorem,
\[
V^T(G_S^T) = \max_{p_T} p_T\left(G_S^T(\overline{v}) - G_S^T(y_S^T(p_T))\right)
\]
is continuous in $G_S^T$,\footnote{If $F_n \to F$ in the weak-$*$ topology, then $F_n(x) \to F(x)$ \textit{only} at all continuity points of $F(\cdot)$. Since we restrict to continuous CDFs, this technical issue does not arise.} and $M_T(G_S^T)$ is non-empty and compact. This establishes the seller's optimal response. 

Furthermore, the period-$T-1$ indifference condition is:
\[
w_{T-1} - p_{T-1} = \mathbb{E}_{\sigma_T}\left[ \delta \left( w_{T-1} - p_T \right) \right],
\]
since at $t = T-1$ there is no ex-ante informational value from delaying. This pins down $w_{T-1}$ uniquely, given anticipated future play, and implies:
\[
p_{T-1} \ge \mathbb{E}_{\sigma_T}[p_T].
\]

\medskip 

\noindent \textbf{The inductive step for Nature.}  
Consider $t = k < T$, assuming the market has not yet cleared, and let $F_S^k$ be the posterior distribution of $v$ among remaining buyers.

\smallskip
\noindent\textit{Case 1: $p_k \le \underline{v}_S^k$.}  
If the seller posts $p_k \le \underline{v}_S^k$, the buyer purchases with probability~1 in equilibrium, regardless of Nature’s choice.

\smallskip
\noindent\textit{Case 2: $p_k > \underline{v}_S^k$.}  
By the induction hypothesis, Nature’s continuation strategy from $t = k+1$ onward employs thresholds $\{w_{k+s}\}_{s \ge 1}$ that yield the same expected payoff to the buyer as if no further information were provided.  
The unique cutoff $w_k$ is therefore determined by:
\[
w_k - p_k = \mathbb{E}_{\sigma_{k+1}}\big[ \delta (w_k - p_{k+1}) \big].
\]
Let Nature’s binary information structure be $(\pi_k, \{\overline{s}_k, \underline{s}_k\})$.  
If $\mathbb{E}_{F_S^k}[v \mid \pi_k, s_k] > w_k$, the buyer strictly prefers to purchase at $t = k$.

We claim: if $\overline{s}_k$ occurs with positive probability, then $\mathbb{E}[v \mid \underline{s}_k] = w_k$.  
Suppose instead $\mathbb{E}[v \mid \underline{s}_k] < w_k$. Modify $\pi_k$ to $(\tilde{\pi}_k, \{\overline{s}_k, \underline{s}_k\})$ by reassigning $\overline{s}_k$ to $\underline{s}_k$ with probability $\varepsilon > 0$.  
For sufficiently small $\varepsilon$, $\mathbb{E}[v \mid \tilde{\pi}_k, \underline{s}_k] < w_k$, so the buyer waits after $\underline{s}_k$. At $t = k+1$, Nature can split $\pi_k^{-1}(\underline{s}_k)$ from the new types $\tilde{\pi}_k^{-1}(\underline{s}_k) \setminus \pi_k^{-1}(\underline{s}_k)$, and revert to the original continuation strategy for the first group.  
The second group no longer buys at $t = k$; since the price path is decreasing by induction, the profit from them is at most $\delta \mathbb{E}_{\sigma_{k+1}}[p_{k+1}]$.  
Because
\[
p_k \ge \mathbb{E}_{\sigma_{k+1}}[p_{k+1}] > \delta \mathbb{E}_{\sigma_{k+1}}[p_{k+1}],
\]
this deviation (not detected by the seller) is strictly profitable for Nature — a contradiction. We also claim $\underline{s}_k$ corresponds exactly to buyers with $v \le y_S^k(w_k)$.  
If not, there exist $v' > v''$ with $v' \in \pi_k^{-1}(\underline{s}_k)$ and $v'' \in \pi_k^{-1}(\overline{s}_k)$.  
Swapping $v'$ and $v''$ leaves $\mathbb{E}[v \mid \overline{s}_k]$ above $w_k$ (so buyer actions unchanged) but lowers $\mathbb{E}[v \mid \underline{s}_k]$ strictly below $w_k$. This reduces to the previous case and Nature can further reduce profit when the seller’s strategy is fixed — again contradicting optimality.

\paragraph{\textit{Fixed-point condition.}}  
Because Nature can lower the threshold slightly to make the buyer’s incentive strict, the equilibrium cutoff must:
\[
w_k - p_k \in \delta \big[ w_k - \bar{m}_S^k(y_S^k(w_k)) \big],
\]
equivalently,
\begin{equation}
p_k = (1-\delta) w_k + \delta \,\bar{m}_S^k\big( y_S^k(w_k) \big),
\label{eqn:fix}
\end{equation}
where $\bar{m}_S^k(\cdot)$ is the convex hull of $m_S^k(\cdot)$, and
\[
m_S^k(b) = \arg\max_{p_{k+1}} \Big\{ p_{k+1} \big( G_S^k(b) - G_S^k(y_S^{k+1}(w_{k+1}(p_{k+1}))) \big) + \delta V^{k+2}(G_S^{y_S^{k+1}, k}(\cdot)) \Big\}.
\]
The objective in $m_S^k$ has strict single-crossing in $(p_{k+1}, b)$ via the term $p_{k+1} G_S^k(b)$, so by the Monotone Selection Theorem \citep{MilgromShannon}, any selection from $m_S^k(b)$ is non-decreasing.  
Since $y_S^k(w_k)$ is continuous and strictly increasing in $w_k$, there exists a unique continuous, non-decreasing function $w_k(p_k)$ satisfying \eqref{eqn:fix}.\footnote{Existence follows from Berge’s Maximum Theorem and upper-hemicontinuity.}  
Thus $w_k$ is uniquely determined by $p_k$. To induce cutoff $w_k = c$, the seller posts $p_k = \max w_k^{-1}(c)$, which maximizes current profit without affecting future play or today’s purchasing set.\footnote{Seller randomization after deviations may still be necessary, as in standard equilibrium constructions.} This eliminates all the randomization on-path, and the indifference condition simplifies to:
\[
w_k - p_k = \mathbb{E}_{\sigma_{k+1}}\big[ \delta (w_k - p_{k+1}) \big]
\quad\Rightarrow\quad
w_k - p_k = \delta (w_k - p_{k+1}),
\]
consistent with \citet{FLT1985} and \citet{GSW}.

\medskip 

\noindent \textbf{The inductive step for the seller.}  
Consider the seller’s pricing decision in period \( t = k \) following the history \( p^{k-1} \).

\smallskip
\noindent\textit{Case 1: Market should have cleared but has not.}  
If, under the equilibrium strategies \( (\Pi, \alpha) \) (allowing for possible seller deviation), the market is predicted to have cleared yet continues, this history is off-path.  
As before, we may assign arbitrary beliefs and allow the seller to post any price that deters such deviations.

\smallskip
\noindent\textit{Case 2: Market not cleared on-path.}  
If the market is not cleared and this is consistent with the equilibrium path, the remaining posterior over \( v \) is \( F_S^k \) with unconditional CDF \( G_S^k \).  
For any \( p_k \) in the support of \( \sigma_k \) we have:
\begin{equation*}
p_k \in M_k(G_S^k)
= \arg\max_{p_k} \left[ p_k \left( G_S^k(\overline{v}) - G_S^k(y_S^k(w_k(p_k))) \right) + \delta V^{k+1}(G_S^{y_S^k,k+1}(\cdot)) \right],
\end{equation*}
and the value function satisfies:
\begin{equation*}
V^k(G_S^k)
= \max_{p_k} \left[ p_k \left( G_S^k(\overline{v}) - G_S^k(y_S^k(w_k(p_k))) \right) + \delta V^{k+1}(G_S^{y_S^k,k+1}(\cdot)) \right].
\end{equation*}
By the induction hypothesis:
\begin{itemize}
    \item \( w_k(p_k) \) is continuous in \( p_k \),
    \item \( y_S^k(w_k) \) is continuous in \( w_k \),
    \item \( G_S^{y,k+1}(\cdot) \) is continuous in \( y \), and
    \item \( V^{k+1}(\cdot) \) is continuous.
\end{itemize}
Therefore the objective
\[
p_k \left( G_S^k(\overline{v}) - G_S^k(y_S^k(w_k(p_k))) \right)
+ \delta V^{k+1}(G_S^{y_S^k,k+1}(\cdot))
\]
is jointly continuous in \( (p_k, G_S^k) \).  
By Berge’s Maximum Theorem, \( V^k(G_S^k) \) is continuous in \( G_S^k \), and \( M_k(G_S^k) \) is non-empty and compact. Because the period-\( k \) information structure leaves the delaying buyer indifferent, her payoff equals that from buying immediately in period \( k \) at price \( p_k \).  
Thus, at \( t = k-1 \) the indifference condition is:
\[
w_{k-1} - p_{k-1}
= \delta \cdot \mathbb{E}_{\sigma_k}[ w_{k-1} - p_k ],
\]
where \( w_{k-1} \) defined this way is the unique cutoff anticipating continuation play.  
It follows immediately that:
\[
p_{k-1} \ge \mathbb{E}_{\sigma_k}[p_k].
\]

\smallskip
Iterating the seller’s inductive step together with the inductive step for Nature from \( t = T \) backward completes the proof.

\item \textbf{Equilibrium strategy profile.}  
The constructed equilibrium has the following on-path strategies:

\begin{itemize}
\item \emph{Deterministic play on path.}  
The strategies of the buyer, the seller, and Nature are deterministic on the equilibrium path, following possible seller randomization in period $t = 1$.

\item \emph{Nature’s strategy.}  
In each period $t$---whether $h_S^t$ is on-path or off-path---Nature reveals to the buyer whether
\[
v \leq y_S^t\big( w_t(p_t) \big).
\]
\item \emph{Seller’s strategy.}  
In period $t$, the seller chooses $p_t$ to solve
\begin{equation*}
\max_{p_t} \;
\sum_{s=t}^T \delta^{\,s-t} \, p_s(p_t) \,
\frac{F\big(y(w_{s-1}(p_t))\big) - F\big(y(w_s(p_t))\big)}{F\big(y(w_{t-1})\big)},
\end{equation*}
where $p_s(p_t)$ denotes the price posted in period $s$ along the continuation path following $p_t$ in period $t$.

\item \emph{Buyer’s strategy.}  In period $t$, the buyer purchases if and only if
\[
v > y_t,
\]
given the signal from Nature; otherwise, she waits.
\end{itemize}
\end{enumerate}
\end{proof}

\begin{proof}[Proof of Proposition \ref{prop:infhorizon}]\
\begin{enumerate}[label=\textbf{Step \arabic*.}, wide=0pt, leftmargin=*, align=left, labelwidth=*, itemsep=1ex, listparindent=\parindent, parsep=0pt]

\item \textbf{Threshold learning.}  
We first show that in any monotone equilibrium, Nature must employ a threshold information structure in every period.
 
Fix an arbitrary monotone equilibrium and define
\[
\lambda_t := \Pr\big[ \text{buyer is recommended to purchase in period $t$} \,\big|\, p^t \big],
\]
the on-path probability—conditional on the realized price history $p^t$—that Nature recommends purchase in period $t$.  
Let $\lambda_\infty$ be the probability that the buyer is never recommended to purchase, and let $y_\infty$ denote the buyer’s expected valuation conditional on this event. We can construct a sequence of price-dependent thresholds
\[
\infty = v_0 \ge v_1 \ge v_2 \ge \cdots \ge v_\infty = 0
\]
such that each $v_t$ depends only on the realized price path $p^t$, and satisfies
\[
\Pr\!\left[ v_t < v \le v_{t-1} \,\middle|\, p^t \right] = \lambda_t
\]
for every $t$ and every price history $p^t$.  
Under this process, in period $t$ Nature recommends “buy” if $v > v_t$ and “wait” if $v \le v_t$.

Consider a deviation where Nature uses this threshold-based rule in every period.  
Since the seller observes only the price path, he cannot detect this deviation, so the price sequence is unchanged.  
For any history $p^t$, a buyer who is recommended to wait now faces a posterior distribution that is (by construction) \emph{inferior in the FOSD sense} to the original equilibrium posterior.  
By the monotonicity property, such buyers will still obey the “wait” recommendation. Under the deviation, for every period $t$ and price history $p^t$, at least as much buyer mass is deferred as in the original equilibrium.  
Hence total social surplus weakly decreases.  
Buyers who are recommended to purchase in a given period retain the option to buy immediately or wait, so their utility does not decline.  
Since buyers’ utilities do not fall and total surplus weakly decreases, the seller’s payoff must fall—implying that Nature’s payoff increases.  
Therefore, in any monotone equilibrium, Nature’s on-path strategy can be taken to be a threshold rule in every period.

\item \textbf{Market clearing in finite time under threshold learning.}  
Using the notation from above, for any $b$ let $F^b$ denote the cdf of the lower part of the prior $F$ truncated at $b$. We first show there exists $b^*$ such that if the seller faces posterior $F^{b^*}$, he optimally posts price $\underline{v}$ and clears the market. Suppose the seller faces $F^b$ at $h_S^t$. If he charges $p_t$, one feasible strategy for Nature is to \emph{fully reveal} the buyer’s type. Then:
\[
V(F^b) \le p_t\big(F(b) - F(w_t)\big) + \delta F(w_t)w_t
\le w_t\big(F(b) - F(w_t)\big) + \delta F(w_t)w_t,
\]
where the second inequality uses $p_t \le w_t$.  
On the other hand, by posting $\underline{v}$ and selling to all remaining buyers,  
\[
V(F^b) \ge F(b)\,\underline{v}.
\]
Note that $V(\cdot)$ is defined relative to a particular equilibrium (since multiple equilibria may exist). Because $F^{-1}$ is Lipschitz-continuous at $0$, there exists $q^*>0$ and $L<\infty$ such that
\[
F^{-1}(q) - \underline{v} \le Lq \quad \forall q \in [0,q^*],
\]
which implies
\[
v - \underline{v} \le L\,F(v) \quad \forall v \in [\underline{v},\,F^{-1}(q^*)].
\]
Take $b \le F^{-1}(q^*)$. Combining the above bounds on $V(F^b)$:
\begin{align*}
0 &\ge F(b)\,\underline{v} - w_t\big(F(b)-F(w_t)\big) - \delta F(w_t)w_t \\
  &\ge F(b)\,\underline{v} - \big(LF(w_t) + \underline{v}\big)\big(F(b)-F(w_t)\big) - \delta F(w_t)\big(LF(w_t) + \underline{v}\big) \\
  &\ge (1-\delta)F(w_t)\,\underline{v} - L F(w_t)\big(F(b)-F(w_t)\big) - \delta L F(w_t)^2 \\
  &\ge F(w_t)\big[ (1-\delta)\underline{v} - L F(b) + L F(w_t) - \delta L F(w_t) \big] \\
  &\ge F(w_t)\big[ (1-\delta)\underline{v} - L F(b) - \delta L F(b) \big].
\end{align*}
The final term is positive for $b$ sufficiently small, which forces $w_t = \underline{v}$ and hence $p_t = \underline{v}$. Thus there exists $b^*$ such that if the seller’s posterior is $F^{b^*}$, he clears the market by setting $p_t = \underline{v}$.

\smallskip
\noindent\textbf{From thresholds to finite-time clearing.}  
Suppose Nature uses a threshold information arrival process on-path with thresholds
\[
\infty = v_0 \ge v_1 \ge v_2 \ge \cdots \ge v_\infty = 0.
\]
By Nature’s sequential rationality, in each $t$ buyers recommended to wait do so. Let $y_t$ be the upper bound of the support of the posterior at time $t$.  
Starting from any $y > b^*$, and for any $\epsilon > 0$, there exists finite $k$ such that $\epsilon$ mass of buyers exit the market within $k$ periods. If not, then:
\[
V(F^y) \le \epsilon\,\overline{v} + \delta^k\overline{v},
\]
which can be made arbitrarily small as $\epsilon \to 0$ and $k \to \infty$. But also
\[
V(F^y) \ge F(y)\,\underline{v} \ge F(b^*)\,\underline{v} > 0,
\]
a contradiction. Therefore, the market must be cleared within
\[
\overline{T}(\delta) = \left\lceil \frac{k(1 - F(b^*))}{\epsilon} + 1 \right\rceil
\]
periods.

\item \textbf{Backward induction.}  
Since the market clears in finite time after any history, we can solve for the equilibrium by backward induction on $t$ and the finite $T$ exactly as in the proof of Theorem~\ref{thm:baseline}. This induction terminates at $T = \overline{T}(\delta)$.
\end{enumerate}
\end{proof}

\begin{proof}[Proof of Theorem \ref{thm:selfconfirming}]\
\begin{enumerate}[label=\textbf{Step \arabic*.}, wide=0pt, leftmargin=*, align=left, labelwidth=*, itemsep=1ex, listparindent=\parindent, parsep=0pt]

\item \textbf{Fixing a price path and admissible threshold processes.}
Fix an arbitrary deterministic price path $(p_1,p_2,\ldots)$. By Proposition~3 in \citet{firstpaper}, the worst‐case information structure against any fixed price path can be taken to be a \emph{threshold} process (not necessarily myopic). Hence Nature’s choice is summarized by a nonincreasing sequence $(y_t)_{t\ge1}$, with the buyer purchasing at the first $t$ such that $v>y_t$.

Under any such process, at time $t$ the IC constraint is
\begin{equation}
\int_{\underline v}^{y_t} (v-p_t) f(v)\,dv
\;\le\;
\sum_{s=t+1}^{\bar T} \delta^{\,s-t}
\int_{y_s}^{y_{s-1}} (v-p_s) f(v)\,dv,
\label{eq:generalindiff}
\end{equation}
and the seller’s discounted profit is
\begin{equation}
\sum_{s=1}^{\bar T} \delta^{\,s-1}
\int_{y_s}^{y_{s-1}} p_s f(v)\,dv.
\label{eq:generalprofit}
\end{equation}
Our goal is to show that for every $t$ either (i) $y_t=y_{t-1}$, or (ii) \eqref{eq:generalindiff} binds at $t$.

\item \textbf{Block decomposition of the price path and Pooling thresholds within a block.}
Let $p_0\equiv+\infty$ and define recursively the block start times
\[
t_1:=1,\qquad
t_k:=\min\{t>t_{k-1}:\ \delta^{\,t-t_{k-1}} p_t < p_{t_{k-1}}\},\quad k\ge2.
\]
Thus, within each block $\{t_k,\ldots,t_{k+1}-1\}$ we have the monotonicity of \emph{discounted} prices:
\begin{equation}
p_{t_k}\ \le\ \delta\,p_{t_k+1}\ \le\ \cdots\ \le\ \delta^{\,t_{k+1}-t_k-1}p_{t_{k+1}-1}.
\label{eq:block-monotone}
\end{equation}
We first prove that for any $k$, it is (weakly) optimal for Nature to \emph{pool} the thresholds inside the block:
\begin{equation}
y_{t_k}=y_{t_k+1}=\cdots=y_{t_{k+1}-1}
\label{eq:block-pool}
\end{equation}
with the pooled level fixed at $y_{t_{k+1}-1}$.

\emph{IC under pooling.}  Start from any profile $\{y_{t_k},y_{t_k+1},\ldots,y_{t_{k+1}-1}\}$ and replace it by \eqref{eq:block-pool} with the pooled level fixed at $y_{t_{k+1}-1}$.  
By \eqref{eq:block-monotone}, for every $v$ and every $\ell\in\{1,\ldots,t_{k+1}-t_k-1\}$,
\[
(v-p_{t_k})\ >\ \delta^{\,\ell}\,(v-p_{t_k+\ell}).
\] 
Hence any type who was recommended to \emph{buy} at some $t_k+\ell$ inside the block ($y_{t_k}\geq v >y_{t_{k+1}-1}$) strictly (weakly) prefers buying already at $t_k$ after pooling; any type with $v\le y_{t_{k+1}-1}$ still learns “wait” at $t_k$ and prefers to wait. Thus recommendations of waiting remain obedient after pooling.

\emph{Profit under pooling.}  Pooling moves all within‐block purchases forward to $t_k$. For any type that originally bought at $t_k+\ell$, the seller’s discounted revenue changes from $\delta^{\,\ell}p_{t_k+\ell}$ to $p_{t_k}$, which weakly \emph{decreases} by \eqref{eq:block-monotone}, strictly if some inequality is strict.  
If the pooling (by raising the value of waiting) causes some types to \emph{defer beyond the block}, profit weakly decreases further because purchases can occur only at block starts $\{t_1,t_2,\ldots\}$, and these satisfy
\[
p_{t_1}\ > \ \delta^{\,t_2-t_1}p_{t_2}\ >\ \delta^{\,t_3-t_1}p_{t_3}\ >\ \cdots,
\]
so shifting a purchase from $t_k$ to a later block $t_{k'}$ lowers discounted revenue.  
Therefore pooling within each block is (weakly) profitable for Nature and (weakly) reduces the seller’s profit. Thus, it is without loss to assume there is no information (as the thresholds are equal) and no sale within each block.

After pooling, Nature’s problem reduces to choosing the block–start thresholds $\{y_{t_k}\}_{k\ge1}$ subject to the (block–level) IC constraints
\begin{equation}
\int_{\underline v}^{y_{t_k}} (v-p_{t_k}) f(v)\,dv
\;\le\;
\sum_{s=k+1}^\infty \delta^{\,t_s-t_k}
\int_{y_{t_s}}^{y_{t_{s-1}}} (v-p_{t_s}) f(v)\,dv,
\label{eq:newcon}
\end{equation}
(where the sum is finite if the horizon is finite).  

Of course, this is slightly more general than required for the proof of Theorem~\ref{thm:selfconfirming}, since the equilibrium pricing path in Theorem~\ref{thm:baseline} automatically satisfies $p_1 > \delta p_2 > \delta p_3 > \dots$.

\item \textbf{Identifying the binding constraints.}
Define  $\overline{v}_{t_k}$ such that
\[
(1 - \delta^{\,t_{k+1}-t_{k}}) \,\overline{v}_{t_{k}}
:= p_{t_{k}} - \delta^{\,t_{k+1}-t_{k}} p_{t_{k+1}} 
\]
and
$\overline{y}_{t_k}$ such that
\[
\mathbb{E}[\,v \mid v \le \overline{y}_{t_k}\,] = \overline{v}_{t_k}.
\]
In other words, $\overline{v}_{t_k}$ is the expected valuation indifferent between buying at $t_k$ at price $p_{t_k}$, and buying at $t_{k+1}$ at price $p_{t_{k+1}}$. Because in Step~1 we established that
\[
p_{t_1} \ > \ \delta^{\,t_2-t_1}p_{t_2} \ > \ \delta^{\,t_3-t_1}p_{t_3} \ > \ \cdots,
\]
it follows that $\overline{v}_{t_k} > 0$. In particular, define $\overline{y}_{T}$ such that $\mathbb{E}[\,v \mid v \le \overline{y}_T\,] = p_T$.

We aim to prove that, Nature's optimal threshold learning process takes the following form: for every $t_k$, either $y_{t_k}=y_{t_{k-1}}$ or $y_{t_k}\geq \overline{y}_{t_k}$. If this is the case, then at any $t_k$, there is no information value in the future, and we must have $y_{t_k}=\overline{y}_{t_k}$ as $y_{t_k}>\overline{y}_{t_k}$ will violate the IC constraint.

\medskip
\noindent\textit{Perturbation idea.}  
We construct a perturbation of $(y_{t_j}, y_{t_k})$ that strictly reduces the seller’s profit if both of the following two conditions holds at $t_k$:  
\[
\text{(a) $y_{t_k}< \overline{y}_{t_k}$, \quad (b) $y_{t_k} < y_{t_{k-1}}$.}
\]
We will use $t_k=T$ as an example, but the same argument works for all $t_k$. Suppose $t_k = T$, and assume all previous IC constraints are slack. Then increasing $y_{t_k}$ reduces the seller’s profit, since some mass of buyers who would have purchased at $t_{k-1}$ delay to $t_k$, and we already know $p_{t_{k-1}} > \delta p_{t_k}$. The increase of $y_{t_k}$ is always feasible while leaving all other thresholds $y_t$ unchanged, until one of the following occurs:
\begin{enumerate}
\item $y_{t_k} = y_{t_{k-1}}$.
\item $y_{t_k}\geq \overline{y}_{t_k}$
\item Some earlier IC constraint binds. 
\end{enumerate}
If either of the first two cases occurs, the claim follows immediately. For the third case, note that because multiple earlier IC constraints could bind simultaneously, we focus on the binding IC with the largest time index, denoted by $t_j$.

\medskip
At $t_j$, the binding IC condition is
\begin{equation}
\int_{\underline v}^{y_{t_j}} (v - p_{t_j}) f(v)\,dv
\;=\;
\sum_{s=j+1}^\infty \delta^{\,t_s - t_j}
\int_{y_{t_s}}^{y_{t_{s-1}}} (v - p_{t_s}) f(v)\,dv.
\label{eq:neweq}
\end{equation}

We now increase $y_{t_k}$ while simultaneously adjusting $y_{t_j}$ so that \eqref{eq:neweq} remains satisfied. Because the IC at $t_j$ stays binding, all earlier IC constraints remain valid: from the buyer’s perspective, there is no information value after $t_j$, since she is indifferent between buying and waiting at $t_j$. Moreover, all later IC constraints remain locally satisfied because $t_j$ was chosen as the largest time index with a binding IC, implying that all subsequent IC are slack.

\medskip
\noindent\textit{Derivative computations.}  
Let $y_{t_j}(y_{t_{k}})$ denote the $t_j$–threshold that satisfies \eqref{eq:neweq} given $y_{t_{k}}$. Differentiate \eqref{eq:neweq} w.r.t.\ $y_{t_{k}}$, holding other $y_{t_s}$ fixed for $s \ne j,k$:

\noindent\emph{From the $y_{t_{k}}$ term on the RHS of \eqref{eq:neweq}:}
\begin{equation}
\label{eq:16}
\delta^{\,t_{k}-t_{j}}
\left(
 -(y_{t_{k}} - p_{t_{k}}) 
\right) f(y_{t_{k}}).
\end{equation}
\emph{From the $y_{t_j}$ term:}  
Move the $y_{t_j}$–integral on the RHS of \eqref{eq:16} to the LHS and differentiate:
\begin{equation}
(y_{t_j} - p_{t_j}) f(y_{t_j})
- \delta^{\,t_{j+1}-t_j} (y_{t_j} - p_{t_{j+1}}) f(y_{t_j})
= (1 - \delta^{\,t_{j+1}-t_j}) (y_{t_j} - \overline{v}_{t_j}) f(y_{t_j}),
\label{eq:yt}
\end{equation}
By the chain rule, keeping \eqref{eq:neweq} binding requires that \eqref{eq:neweq} equals \eqref{eq:yt} multiplied by $y_{t_j}'(y_{t_k})$:
\begin{equation}
\delta^{\,t_{k}-t_j}  (p_{t_{k}} - y_{t_{k}}) f(y_{t_{k}})
= (1 - \delta^{\,t_{j+1}-t_j}) (y_{t_j} - \overline{v}_{t_j}) f(y_{t_j}) \, y_{t_j}'(y_{t_{k}}).
\label{eq:indiff}
\end{equation}

We claim that $y_{t_j} > \overline{v}_{t_j}$.  
Indeed, by hypothesis the IC at $t_j$ binds. Recall that $\overline{v}_{t_j}$ is defined as the expected value at which the buyer is indifferent between buying and waiting at $t_j$ even if no future information arrives. If instead $y_{t_j} \leq \overline{v}_{t_j}$, then any buyer told $v \leq y_{t_j}$ would strictly prefer to wait at least until $t_{j+1}$, contradicting the binding of the IC at $t_j$.

This establishes the derivative relationship \eqref{eq:indiff} with $(y_{t_j} - \overline{v}_{t_j}) > 0$, which is the key ingredient for the profit‐sign calculation in the next step.

\medskip
\textit{Effect of the perturbation on profit.}  
We now differentiate the seller’s profit \eqref{eq:generalprofit} under the perturbation of $(y_{t_j},y_{t_{k}})$ constructed above. Since all other thresholds are fixed, only the $t_j$–, $t_{j+1}$–, and $t_{k}$–terms in \eqref{eq:generalprofit} vary. Thus it suffices to differentiate
\begin{align*}
&\quad p_{t_j}\big(1 - F(y_{t_j}(y_{t_{k}}))\big)
+ \delta^{\,t_{j+1}-t_j} p_{t_{j+1}} F(y_{t_j}(y_{t_{k}})) - \delta^{t_k-t_j}p_{t_k}F(y_{t_k})
\end{align*}
where we have made explicit the dependence of $y_{t_j}$ on $y_{t_{k}}$. Differentiating term–by–term with respect to $y_{t_{k}}$ yields:
\begin{align*}
&-p_{t_j} f(y_{t_j}(y_{t_{k}}))\, y_{t_j}'(y_{t_{k}})
+ \delta^{\,t_{j+1}-t_j} p_{t_{j+1}}
 f(y_{t_j}(y_{t_{k}}))\,y_{t_j}'(y_{t_{k}}))-
 \delta^{t_k-t_j}p_{t_k}f(y_{t_k})
\end{align*}
Multiply through by $(y_{t_j} - \overline{v}_{t_j})>0$ (recall this sign was established above) and use \eqref{eq:indiff} to substitute for $y_{t_j}'(y_{t_{k}})$ wherever it appears. After straightforward algebra and factoring out common positive terms, the derivative of profit with respect to $y_{t_{k+1}}$ is proportional to
\begin{equation*}
\big(-p_{t_j} + \delta^{\,t_{j+1}-t_j} p_{t_{j+1}}\big)
\frac{\delta^{\,t_{k}-t_j} }{1 - \delta^{\,t_{j+1}-t_j}}
\big(p_{t_{k}} - y_{t_{k}}\big)
 - \delta^{\,t_{k}-t_j} p_{t_{k}} (y_{t_j} - \overline{v}_{t_j}).
\end{equation*}
Divide by $\delta^{\,t_{k}-t_j}$ and substitute the definitions of $\overline{v}_{t_k}$. The change in profit from increasing $y_{t_{k}}$ is proportional to
\begin{equation}
\overline{v}_{t_j} \big( y_{t_{k}} - p_{t_{k}} \big)
+ p_{t_k} \big( \overline{v}_{t_j} - y_{t_j} \big)
= \overline{v}_{t_j} y_{t_{k}} - p_{t_k} y_{t_j}.
\label{eq:18}
\end{equation}

\medskip
\noindent\textit{Case analysis.}  

\underline{Case 1: $p_{t_k} > \overline{v}_{t_j}$.}  
Since $y_{t_j} \ge y_{t_{k}}$, \eqref{eq:18} is strictly negative. Profit is reduced by increasing $y_{t_{k}}$. This is also slightly more general than required for the proof of Theorem~\ref{thm:selfconfirming}, since the equilibrium pricing path in Theorem~\ref{thm:baseline} automatically satisfies $p_{t_k} <\overline{v}_{t_j}$.

\underline{Case 2: $p_{t_k} \leq\overline{v}_{t_j}$.}  
We want to prove 
\[
\overline{v}_{t_j} y_{t_{k}} - p_{t_k} y_{t_j}
\]
is globally negative. Threshold–ratio monotonicity says $v \mapsto v / y(v)$ is decreasing, so $\frac{p_{t_{k}}}{\overline{y}_{t_{k}}} \ge \frac{\overline{v}_{t_j}}{\overline{y}_{t_j}}$, which implies
\[
\overline{v}_{t_j} \,\overline{y}_{t_{k}} - p_{t_{k}} \,\overline{y}_{t_j} \le 0.
\]
Note we must have $y_{t_j} \geq \overline{y}_{t_j}$. Since the IC at $t_j$ is binding, if $y_{t_j} < \overline{y}_{t_j}$, then any buyer recommended to wait would strictly prefer waiting even in the absence of future information—a contradiction.  

Thus, we conclude globally that (given $y_{t_k}<\overline{y}_{t_k}$)
\[
\overline{v}_{t_j}\,y_{t_{k}} - p_{t_{k}}\,y_{t_j} < 0,
\]
and it is optimal for Nature to increase $y_{t_k}$ while adjusting $y_{t_j}$ until one of the following occurs:
\begin{enumerate}
\item $y_{t_k} = y_{t_{k-1}}$;
\item $y_{t_k}\geq \overline{y}_{t_k}$;
\item $y_{t_j} = y_{t_{j+1}}$;
\item Some constraint between $t_j$ and $t_k$ binds. 
\end{enumerate}
If either of the first two cases occurs, the claim follows. The third case is simply a special instance of the fourth: it means we can identify a larger time index and repeat the same procedure. Since the game has only finitely many periods, and the largest binding constraint can only shift forward in time, eventually we must reach a point at which either of the first two cases occurs.

For $t_k\neq T$, the argument is essentially the same as before except at the last period $T$, adjusting $y_T$ has no effect on future profits (as $T$ is the last period), while in earlier periods we must adjust for the impact on future profits. The logic, however, is unchanged.

By similar algebra, the change in profit from increasing $y_{t_k}$ is proportional to
\[
\overline{v}_{t_j} y_{t_{k}} - \overline{v}_{t_{k}} y_{t_j}.
\]
By threshold–ratio monotonicity, the map $v \mapsto v / y(v)$ is decreasing, so
\[
\frac{\overline{v}_{t_{k}}}{\overline{y}_{t_{k}}} \;\ge\; \frac{\overline{v}_{t_j}}{\overline{y}_{t_j}},
\]
which implies
\[
\overline{v}_{t_j}\,\overline{y}_{t_{k}} - \overline{v}_{t_{k}}\,\overline{y}_{t_j} \leq 0.
\]
Similarly, we must have $y_{t_j} \geq \overline{y}_{t_j}$. Since the IC at $t_j$ is binding, if $y_{t_j} < \overline{y}_{t_j}$, then any buyer recommended to wait would strictly prefer waiting even without future information—a contradiction. Thus, we conclude globally that (given $y_{t_k}< \overline{y}_{t_k}$)
\[
\overline{v}_{t_j}\,y_{t_{k}} - \overline{v}_{t_{k}}\,y_{t_j} < 0,
\]
and it is optimal for Nature to increase $y_{t_k}$ while adjusting $y_{t_j}$.
\end{enumerate}
\end{proof}

\begin{proof}[Proof of Corollary \ref{cor:unidete}]Note that for any given deterministic price path, Theorem~\ref{thm:selfconfirming} and threshold–ratio monotonicity imply that the worst–case learning process is the myopic threshold process. We claim that Nature can implement this process \emph{sequentially}, without committing ex ante: in each period $t_k$, by construction
\[
\mathbb{E}[\,v \mid v \leq y_{t_k}\,] \;=\; \overline{v}_{t_k},
\]
and the buyer is weakly indifferent between purchasing at $t_k$ and waiting until $t_{k+1}$ even if she conjectures that no further information will arrive.

Thus, for any deterministic price path generated by an equilibrium strategy, Nature would deviate to the myopic threshold process, and hence in any deterministic equilibrium Nature uses the myopic threshold process on path. Under this condition, solving for equilibrium reduces to determining the seller’s optimal deterministic price path subject to sequential rationality. The remaining steps follow exactly as in the previous proof.
\end{proof}

\begin{proof}[Proof of Proposition \ref{prop:2}]
When Nature implements myopic threshold learning process, the seller’s discounted expected profit from time $t$ onward (taking $w_{0} = y^{*}(w_{0}) = \overline{v}$) is:
\begin{equation*}
\sum_{s=t}^{T} \delta^{\,s-t} p_{s} \,
\frac{F\!\left(y^{*}(w_{s-1})\right) - F\!\left(y^{*}(w_{s})\right)}{F\!\left(y^{*}(w_{t-1})\right)}.
\end{equation*}
From \cite{GSW} and \cite{AusubelDeneckere1989}, we know there exists an weak-Markov equilibrium which pin downs the seller's equilibrium price path. And Nature and the buyer will have no incentive to deviate just as the proof of Theorem \ref{thm:baseline}.
\end{proof}

\begin{proof}[Proof of Proposition \ref{prop:nogapunobs}]
The argument is exactly the same as in the proof of Corollary \ref{cor:unidete}.
\end{proof}

\subsection{Distributional Assumptions Yielding Threshold-Ratio Monotonicity}

\begin{proof}[Proof of Proposition \ref{prop:suffcond}]

Our goal is to show that $\frac{w}{y(w)}$ is decreasing in $w$, or decreasing in $y(w)$ since it increases in $w$. Let $y = y(w)$, then $w = \mathbb{E}[v \mid v \leq y] = \frac{\int_{v\leq y} vf(v)~dv}{F(y)}$ so that 
\[
\frac{w}{y} = \frac{\int_{v\leq y} vf(v)~dv}{yF(y)}.
\]
The derivative with respect to $y$ is 
\[
\frac{\partial (w/y)}{\partial y} = \frac{yf(y) \cdot yF(y) - (yf(y)+F(y))\cdot ( \int_{v\leq y}vf(v)~dv)}{y^2 F(y)^2}.
\]
Rearranging, this derivative is non-positive if and only if 
\[
\int_{v\leq y} vf(v) ~dv \geq \frac{y^2f(y)F(y)}{yf(y)+F(y)}. 
\]
The above inequality holds at $y = \underline{v}$, so a sufficient condition for it to hold at every $y$ is that the derivatives of two sides are ordered. That is, we want
\[
y f(y) \geq \left(\frac{y^2f(y)F(y)}{yf(y)+F(y)}\right)'. 
\]
We can compute the derivative of $\frac{y^2f(y)F(y)}{yf(y)+F(y)}$ to be 
\[
\frac{(yf(y)+F(y))\cdot (2yf(y)F(y) + y^2f'(y)F(y)+y^2f(y)^2) - y^2f(y)F(y) \cdot (2f(y)+yf'(y))}{(yf(y)+F(y))^2},
\]
which simplifies to
\[
\frac{y^3f(y)^3+y^2f(y)^2F(y) + y^2f'(y)F(y)^2 + 2yf(y)F(y)^2}{(yf(y)+F(y))^2}.
\]
This expression is smaller than $yf(y)$ if and only if
\[
yf(y) (yf(y)+F(y))^2 \geq y^3f(y)^3+y^2f(y)^2F(y) + y^2f'(y)F(y)^2 + 2yf(y)F(y)^2. 
\]
After some more algebra, the desired inequality becomes
\[
y^2f(y)^2F(y) \geq y^2 f'(y)F(y)^2 + yf(y)F(y)^2.
\]
Dividing both sides by $yF(y)$, this is equivalent to
\[
yf(y)^2 \geq yf'(y)F(y) + f(y)F(y). 
\]
We can further divide both sides by $F(y)^2$ to arrive at
\[
y\frac{f(y)^2}{F(y)^2} \geq y\frac{f'(y)}{F(y)} + \frac{f(y)}{F(y)}. 
\]
Let $h(y) = \frac{f(y)}{F(y)}$ with $h'(y) = \frac{f'(y)}{F(y)} - \frac{f(y)^2}{F(y)^2}$. The above inequality then becomes
\[
yh'(y) + h(y) \leq 0. 
\]
Note that $yh'(y) + h(y)$ is the derivative of $yh(y)$, so this reduces to $yh(y)$ decreasing in $y$. \end{proof}

\subsection{Using Learning to Sustain Constant Price Paths and Equilibrium Multiplicity}
\begin{proof}[Proof of Proposition \ref{prop:unboundedTime}]
We consider two cases for this proof: first the case where $T = \infty$, and then the modified argument for $T < \infty$. In both cases, we construct the following equilibrium:

\begin{itemize} 
\item On-path, the seller posts a price equal to the buyer's expected value $\mathbb{E}_F[v]$, and no information is revealed.

\item The buyer randomizes purchase with a probability to be specified—specifically, chosen so that the seller has incentives to follow the equilibrium strategy. 

\item If the seller deviates, the equilibrium reverts to the sequentially worst-case outcome described in Theorem \ref{thm:baseline}/Proposition \ref{prop:infhorizon}.
\end{itemize}

We now prove this profile forms an equilibrium. It is immediate that following a deviation by the seller, the future play constitutes an equilibrium, by Theorem \ref{thm:baseline}/Proposition \ref{prop:infhorizon}. The same holds on-path: since the buyer's purchasing decision does not depend on $v$, the on-path distribution of $v$ conditional on not having purchased at time $t$ is simply $F$. Thus, the buyer is indifferent between purchasing and delaying, as both deliver payoff $0$, making them willing to randomize. Moreover, because we assume the buyer (strictly) randomizes, no profitable deviation is available to them, as all actions occur with positive probability on-path. 

It remains to show that the seller does not prefer to deviate on-path, for appropriately chosen randomization probabilities. Let $r^{*}$ denote the profit obtained in the equilibrium of Proposition \ref{prop:infhorizon}/Theorem \ref{thm:baseline}. The seller obtains at most $r^{*}$ following any deviation; in particular, since the buyer's posterior distribution on-path is always $F$, and the horizon is infinite, this property holds at every time. Suppose we seek an equilibrium where the seller's continuation value is $v^{*}$ at every point in time, with $v^{*} > r^{*}$. In this case, set the buyer's purchase probability to be $\rho$ in every period, where $\rho$ satisfies
\begin{equation*} 
v^{*} = \rho \E_{F}[v] + (1- \rho)\delta v^{*} \quad \Rightarrow \quad 
\rho = \frac{v^{*}(1-\delta)}{\E_{F}[v] - \delta v^{*}},
\end{equation*}
with $\rho \in (0,1)$ whenever $v^{*} \in (r^{*}, \E_{F}[v])$.

Thus, by charging $\E_{F}[v]$, the seller obtains a higher payoff than from deviating. This verifies the conditions in the proposition: (i) the seller uses a constant price path; (ii) the profit obtained is any $v^{*} \in (r^{*}, \E_{F}[v])$; and (iii) the market does not clear in any finite time, since $\rho$ is constant and hence the probability the buyer has not purchased at or before time $K$ is $(1-\rho)^{K} > 0$.

For the $T < \infty$ case, define $v_{T} = \E_{F}[v]$ so that the buyer buys with probability $1$ in the last period. Given any $v_{t+1}$ with $t < T$, define $v_{t}$ and $\rho_{t}$ by
\begin{equation*} 
v_{t} = \rho_{t}\E_{F}[v] + (1- \rho_{t})\delta v_{t+1}. 
\end{equation*}
Let $r_{t}$ denote the $T = t$ equilibrium payoff identified in Theorem \ref{thm:baseline} with prior $F$, clearly $r_t< \E_{F}[v]$ unless it is optimal for the seller to clear the market at price $\underline{v}$ at $t=1$. Hence, the equilibrium can take the same form as above, provided the sequence $v_{1},\rho_{1}, v_{2}, \rho_{2}, \ldots, v_{t-1}, \rho_{t-1}$ (with $v_{T} = \E_{F}[v]$) is such that $v_{t} \geq r_{t}$ for all $t$. This can be done by carefully choosing $\rho_t$ close to $1$ so each $v_t$ is close enough to $\mathbb{E}_F[v]$. In this case, the seller obtains a higher payoff under the constant price path than from deviating, and the buyer remains indifferent between purchasing at any time and thus is willing to follow the mixed strategy.
\end{proof}

\section{Dynamically Inconsistent Informationally Robust Objectives} \label{sect:alternatives}
As we hope the analysis in this paper will be useful more broadly beyond pricing applications, it is instructive to discuss which alternative assumptions we could have adopted. This detour aims to deepen appreciation for our main benchmark while clarifying challenges that may arise in future work. We articulate alternative benchmarks and explain why these are less appealing in the informationally robust dynamic durable goods setting. Of course, this conclusion may not hold in other applications, so it is worth highlighting what some alternatives could be. 

Fully developing each benchmark formally would take us too far afield; instead, we rely on examples or simplifications to clarify how each would have affected the analysis, thereby providing intuition for the impact of our modeling choices. Throughout this section, we again focus exclusively on the gap case, while fully maintaining the basic structure of the game; therefore, Sections \ref{subsect:underlying}, \ref{subsect:timing}, and \ref{subsect:strategies} apply in their entirety. Instead, we consider alternative solution concepts distinct from Definition \ref{def:wctcc}.

In all three cases we discuss, the seller is assumed to set prices each period under the belief that Nature has \textit{committed} to arbitrary learning processes and that the buyer has already observed the learning process—in contrast to our baseline model, where Nature commits only within period. These cases differ along two dimensions:

\begin{enumerate}
\item Whether the seller chooses prices anticipating that the worst case will change over time. This determines how the seller sets today’s price.
\end{enumerate} 

\noindent Section \ref{sect:naive} considers the naive case, where the seller does not take such changes into account, and thus assumes the $\tilde{t}$ price would be part of an equilibrium outcome under the worst-case learning process at time $t$, for all $\tilde{t} > t$. Sections \ref{sect:sophistication} and \ref{sect:worsepast} instead consider sophisticated sellers, who recognize that the time-$\tilde{t}$ price will be optimal against a different learning process than the worst case at time $t$.

\begin{enumerate} 
\setcounter{enumi}{1}
\item Whether the worst-case learning process at time $t > 1$ is restricted to the information structures Nature would have chosen in equilibrium at time $\tilde{t} < t$.
\end{enumerate} 

\noindent  As mentioned in the main text, the sequentially worst case implicitly assumes that the seller does not revisit ``worst-case'' scenarios from earlier periods. Section \ref{sect:worsepast} considers the case where Nature can re-optimize over past information structures; in Sections \ref{sect:naive} and \ref{sect:sophistication}, it cannot.

\medskip

\noindent Note that none of these three cases can generally be formulated as saddle points of a game between the seller and Nature. They are therefore less grounded in game theory and less directly related to the standard robustness approach.

\subsection{Naivet\'e over Future Actions} \label{sect:naive}
An alternative would be to assume that the seller \emph{does} consider the worst case over all learning processes, but fails to recognize that this worst case will change over time, and thus does not anticipate that his future choices will differ. Under this assumption, the seller displays na\"{i}vet\'{e}: he simply expects himself to take certain actions in the future and considers a worst case with respect to those actions, failing to recognize that the worst case changes with $t$. 

Specifically, suppose that at every time $t$, for $t=1,2,\ldots$, the seller chooses $p_{t}$ by considering the following game:
\begin{enumerate}
\item Nature \textbf{commits} to a strategy
\[
\Pi : \bigcup_{t} H_{N}^{t} \to \Delta\left(\left\{(\pi,S)\right\}\right),
\]
\item The seller and the buyer play the dynamic game taken Nature's strategy as given. The seller chooses $\sigma : \cup_{t} H_{S}^{t} \rightarrow  \Delta(\R_{+})$.
The buyer chooses $\alpha$, given $\mu$ \textbf{and $\Pi$}, for all $h_{B}^{t}$, maximizing the continuation payoff conditional on reaching $h_{B}^{t}$:
\[
\mathbb{E}_{\mu,\mathbb{P}_{\sigma,\alpha,\Pi}}\!\left[\sum_{\tau \ge t} \delta^{\tau-t} (v-p_{\tau})\,\mathbf{1}_{\{\text{accept at }\tau\}} \,\big|\, h_{B}^{t}\right],
\]
where $\tau$ is the induced stopping time.
\item Nature chooses strategy minimizing the seller's payoff in the equilibrium induced. 
\end{enumerate}
Note that because the seller and the buyer obverse Nature's strategy (unlike in the baseline model), Nature can be viewed as having greater ``commitment power''. To see what this  ``commitment power'' implies, given strategies $(\sigma,\alpha,\Pi)$, denote the seller's expected surplus starting from $t=2$ by $R_{2}$. Then the seller's expected surplus is
\[
mp_{1} + \delta (1-m)R_{2},
\]
where $m$ denotes the mass of buyers who purchase at $t=1$. Now suppose Nature modifies the original $t=2$ strategy $\Pi(h_{N}^{2})$ as follows:
\begin{enumerate}
\item If the seller charges $R_{2}$, Nature provides no information.
\item If the seller charges any other price, Nature implements the original strategy.
\end{enumerate}
It is then optimal for the seller to charge $R_{2}$ at $t=2$ given $\sigma$ is sequentially rational. The market clears in the second period, because $R_{2}$ is the seller's surplus, which is strictly less than the total surplus $\E[v \mid h_{S}^{2}]$. Since no further information arrives, all remaining buyers purchase at $t=2$.

We now prove that the buyers who originally waited at $t=1$ strictly prefer to wait under this modification. Clearing the market at $t=2$ maximizes total surplus, since additional delay only decreases total surplus. Because the seller’s expected surplus remains unchanged, the waiting buyer’s expected surplus must increase, making buyers more willing to wait at $t=1$.

Of course, some buyers who originally purchased at $t=1$ may now delay to $t=2$, but this only hurts the seller: further delay reduces total surplus, while buyer surplus cannot fall since buyers are always optimizing.

Thus, we have shown that for any learning process and the seller--buyer equilibrium induced by the dynamic game, there exists another learning process of the form above that induces an equilibrium in which the seller obtains a lower profit. Thus, the seller-worst equilibrium must take the above form. Consider the following example, which illustrates the solution the seller anticipates when choosing his first-period price:

\begin{example} \label{worseinfoexample}
Suppose $T=2$ and $F\sim U[0,2]$, consider the following learning process: 

\begin{itemize} 
\item In the first period, the seller charges some $p_{1}^{*}$ on-path; following any $p_{1}$, the buyer learns whether $v > \tilde{v}(p_{1})$ and buys if and only if it is. We leave $\tilde{v}(p_{1})$ as to-be-specified for now. 

\item In the second period, the seller charges price $\frac{\tilde{v}(p_{1})}{8}$, the buyer receives no additional information, and the buyer purchases.
\item If the seller deviates in the second period to a price $\hat{p}_{2} \neq \frac{\tilde{v}(p_{1})}{8}$, the buyer learns whether or not $v > 2\hat{p}_{2}$. 
\end{itemize}
Since all remaining buyers in the second period have $v \leq \tilde{v}(p_{1})$, the above construction ensures that the seller has no (strictly) profitable second-period deviation following any first period price. Indeed, in $t=2$, on-path the seller obtains profit $\frac{\tilde{v}(p_{1})}{8} \cdot \frac{\tilde{v}(p_{1})}{2}$, where $\frac{\tilde{v}(p_{1})}{8}$ is the price and $\frac{\tilde{v}(p_{1})}{2}$ is the probability that $v \leq \tilde{v}(p_{1})$. But as shown in the main text, the best alternative price $p_2$ for the seller is $p_2 = \frac{\tilde{v}(p_{1})}{4}$, which delivers the same profit level $\frac{\tilde{v}(p_{1})}{4} \cdot \frac{\tilde{v}(p_{1})}{4}$ where $\frac{\tilde{v}(p_{1})}{4}$ is both the price and the probability that $v \leq \frac{\tilde{v}(p_{1})}{2}$, the threshold that Nature would use. Since $\E[v \mid v < \tilde{v}]=\tilde{v}/2$, if the buyer learns that $v < \tilde{v}$ in period 1 and does not buy, then the buyer obtains $\frac{3 \tilde{v}}{8}$ in the second period. Since \emph{every} buyer with $v\le\tilde{v}$ faces the same information set (and in particular, chooses the same action), the value of $\tilde{v}$ such that the buyer is indifferent between buying at time 1 and delaying purchase to time 2 satisfies:  
\begin{equation*} 
\frac{\tilde{v}}{2} - p_{1} = \delta \frac{3 \tilde{v}}{8} \Rightarrow  \tilde{v} = \frac{8p_{1}}{4- 3 \delta}. 
\end{equation*}
Suppose that Nature, in the first period, tells the buyer whether her value is above or below $\frac{8p_{1}}{4-3\delta}$. Given this information structure (as well as understanding that the seller will follow the equilibrium strategy), the buyer will delay if told her value is below the threshold and not if it is above the threshold. Since the probability the buyer's value is above the first period threshold is $1 - \frac{4p_{1}}{4-3\delta}$ (since $v \sim U[0,2]$), the seller's profit can be written as: 
\begin{equation*} 
p_{1} \left(1 - \frac{4p_{1}}{4-3\delta} \right) + \delta  \frac{4p_{1}}{4- 3 \delta}  \frac{p_{1}}{4 - 3 \delta}  
\end{equation*}
Take first order condition:
\[
1 - \frac{8 p_{1}}{4- 3 \delta}+ \frac{8 p_{1} \delta}{(4-3 \delta)^{2}} = 0 \Rightarrow p_{1} = \frac{(4-3\delta)^{2}}{32(1-\delta)}.
\]
Profit at this price is: 
\begin{equation*} 
\frac{(4-3\delta)^{2}}{32(1-\delta)} \left(1- \frac{4(4-3\delta)}{32(1-\delta)} \right) + \delta \frac{4(4- 3\delta)^{2}}{(32(1- \delta))^{2}}= \frac{(4- 3\delta)^{2}(32(1- \delta)-4(4-3\delta)+4 \delta)}{(32(1- \delta))^{2}}= \boxed{\frac{(4-3\delta)^{2}}{64(1- \delta)}}
\end{equation*}
We check that this solution does indeed involve interior solution so first order condition is sufficient. Given $p_{1}$, we have $\tilde{v}=2$ if: 
\begin{equation*} 
1- \frac{(4-3\delta)^{2}}{32(1-\delta)}= \delta \frac{3}{4} \Rightarrow \delta=4/5. 
\end{equation*}
So, if $\delta < 4/5$, this scheme involves profit exactly as above. If $\delta  \geq 4/5$, all buyers delay to the second period and no sale occurs in the first period, meaning the total profit is $\delta/4$. 
The optimal price in the first period given $\tilde{v}(p_1)$ is chosen to minimize the seller's profit is $p_1=\frac{(4-3\delta)^2}{32(1-\delta)}$. 
\end{example}
This example illustrates what the seller will ``think'' and how he will set the first-period price $p_1$. In the second period, however, providing no information to the buyer is not worst case. Thus, instead of charging $\tilde{v}(p_1)/8$ as in Section 3.2, the seller will charge $\tilde{v}(p_1)/4$. Once the second period begins, the seller updates his conjecture (realizing that Nature will not implement the commitment solution in the first period) and therefore charges the optimal price in this subgame, which is precisely $\tilde{v}(p_1)/4$.

One important distinction is that this example does not contradict Theorem \ref{thm:selfconfirming}, even if the prior $F$ is still the uniform distribution. The reason is that in this example the seller and buyer act after observing Nature's strategy, so the seller's pricing strategy is endogenous rather than exogenous.

\begin{example} \label{ex:weirdthing}
Take $T = \infty$ and $v \sim U[0,2]$. The sequentially worst-case equilibrium outcome with $v \sim U[0,2]$ coincides with the known-values case with $v \sim U[0,1]$. The Coasian equilibrium with $v \sim U[0,1]$ is solved in \cite{GSW} and \cite{Stokey81}.\footnote{The known-values case has a unique outcome, for fixed $\delta$, when $v \sim U[\varepsilon,1]$, which converges to the Coasian outcome as $\varepsilon \to 0$. For our purposes, the same point can be made by considering a sufficiently small $\varepsilon$.} In the known-values case with $v \sim U[0,1]$, the seller's profit when $\tilde{v}$ is the highest buyer value remaining is
\begin{equation*} 
r^{*}(\tilde{v}) = \frac{1}{2}\left(1 - \frac{1}{\delta} + \frac{1}{\delta}\sqrt{1 - \delta}\right)\tilde{v}^{2}.
\end{equation*}

\noindent One can verify that $\lim_{\delta \to 1}r^{*}(1) = 0$, as predicted by the Coase conjecture.

We now guess and verify a naivete equilibrium. In this equilibrium, 
\begin{enumerate}
\item At $t=1$, the seller charges $p_1$, and threshold $\tilde{v}$ is revealed.
\item At $t=2$, the seller charges $p_2$, all buyers purchase, and Nature provides no further information.
\end{enumerate}

Since $\tilde{v}$ must make the buyer indifferent between purchasing and not when learning $v < \tilde{v}$, we have
\[
\frac{\tilde{v}}{2} - p_1 = \delta\left(\frac{\tilde{v}}{2} - p_2\right).
\]
The implied profit is
\begin{equation*} 
p_{1}\left(1 - \frac{\tilde{v}}{\bar{v}}\right) + \delta \frac{\tilde{v}}{\bar{v}} p_2.
\end{equation*}
By uniformity, price should be linear in $\bar{v}$. Thus, suppose the seller's (optimal) prices are $p_1 = k_1 \bar{v}$ and $p_2 = k_2 \tilde{v}$. By the indifference condition, we obtain
\[
\tilde{v} = \frac{k_1}{\tfrac{1}{2} - \tfrac{\delta}{2} + \delta k_2}\bar{v}.
\]
The profit is therefore
\[
\bar{v}\Bigg[k_1\left(1 - \frac{k_1}{\tfrac{1}{2}-\tfrac{\delta}{2}+\delta k_2}\right) 
+ \delta k_2 \left(\frac{k_1}{\tfrac{1}{2}-\tfrac{\delta}{2}+\delta k_2}\right)^2\Bigg].
\]
Now, the seller expects to be able to capture, in the second period, exactly the continuation surplus that remains after the first period, according to the previous construction.  In the second period, the seller must charge a price equal to the available surplus, implying
\[
k_2 \tilde{v}\frac{\tilde{v}}{\bar{v}} =
\tilde{v}\Bigg[k_1\left(1 - \frac{k_1}{\tfrac{1}{2}-\tfrac{\delta}{2}+\delta k_2}\right) 
+ \delta k_2 \left(\frac{k_1}{\tfrac{1}{2}-\tfrac{\delta}{2}+\delta k_2}\right)^2\Bigg].
\]
Solving yields
\[
k_2\Big(\tfrac{1}{2}-\tfrac{\delta}{2}+\delta k_2\Big) =
\Big(\tfrac{1}{2}-\tfrac{\delta}{2}+\delta k_2\Big)^2 - k_1\Big(\tfrac{1}{2}-\tfrac{\delta}{2}+\delta k_2\Big) + \delta k_1 k_2,
\]
so that
\[
k_1 = \tfrac{1}{2}\big(1 - 2k_2 - \delta + 4k_2\delta - 4k_2^2\delta\big).
\]
Now suppose in the first period the seller charges $k \bar{v}$. Then we must have
\[
\frac{\tilde{v}}{2} - k\bar{v} = \delta\left(\frac{\tilde{v}}{2} - k_2\tilde{v}\right).
\]
Solving gives 
\[
\tilde{v} = \frac{k}{\tfrac{1}{2}-\tfrac{\delta}{2}+\delta k_2}\bar{v}.
\]
The corresponding profit is
\[
k\bar{v}\left(1 - \frac{k}{\tfrac{1}{2}-\tfrac{\delta}{2}+\delta k_2}\right) 
+ \delta \frac{k}{\tfrac{1}{2}-\tfrac{\delta}{2}+\delta k_2}\cdot \frac{kk_2}{\tfrac{1}{2}-\tfrac{\delta}{2}+\delta k_2}\bar{v}.
\]
This uses the one-shot deviation principle to determine how $p_2$ changes with $p_1$. Taking the first-order condition with respect to $k$, we have
\[
k^* = \frac{\big(\tfrac{1}{2}-\tfrac{\delta}{2}+\delta k_2\big)^2}{1-\delta}.
\]
Because $k^* = k_1$ in equilibrium, we require
\[
\frac{\big(\tfrac{1}{2}-\tfrac{\delta}{2}+\delta k_2\big)^2}{1-\delta}
= \tfrac{1}{2}\big(1 - 2k_2 - \delta + 4k_2\delta - 4k_2^2\delta\big).
\]
Solving yields
\[
k_2 = \frac{1-\delta}{4-2\delta}, 
\quad k_1 = \frac{1-\delta}{(2-\delta)^2}.
\]
Thus the seller first period price is $\frac{1-\delta}{(2-\delta)^2}\bar{v}$, and $\tilde{v}=\frac{4}{(2-\delta)^3}\bar{v}$. Note that indeed when $\delta=0$, this reduces to one period static case, as expected.

\end{example}
Here, when $\delta \geq 2 - \sqrt[3]{4} \approx 0.4126$, the seller again does not attempt to sell in the first period. A key difference, however, is that the horizon is now infinite. As a result, the seller's problem at time 2 \emph{looks identical} to the problem at time 1 whenever sale occurs with probability 0 in period 1.

This observation shows that \emph{the seller would never induce a sale} in this alternative, for this specification with sufficiently high $\delta$—and, importantly, $\delta$ need not be particularly close to one for this to occur. After waiting one period, the seller effectively ``resets'' the worst case. This property is unusual and highlights how, in principle, the use of the maxmin objective can \emph{dramatically change} the pricing strategies a seller might adopt. We are not aware of other environments where the seller does not even \emph{attempt} to sell in equilibrium. 

On the other hand, this result also provides a reason why our benchmark may be more useful than the fully-pessimal-and-naive case. It seems difficult to imagine that a seller, capable of computing discounted payoffs, would not anticipate never even attempting to sell under this objective. We note that a similar phenomenon can also arise in Bayesian models with a finite horizon (e.g., \cite{Fershtman1993}).

\subsection{Sophistication} \label{sect:sophistication}
While the previous section shows that the worst-case information structure for the seller at $t=1$ will generally induce an equilibrium where the seller does not optimize against the worst case at $t=2$, one might instead insist that the seller \emph{maximizes} against the worst-case learning process, while \emph{acknowledging} that this may change over time. Such a seller is dynamically inconsistent, but aware of this fact. 

To be precise, this alternative induces the following assumptions regarding the objectives of each player:

\begin{itemize} 
\item At $t=1$, the seller chooses $p_{1}$ anticipating the equilibrium strategies $\sigma_2(h^2_S)$ he would adopt at $t=2$. In particular, $p_{1}$ is chosen to maximize profit in the equilibrium induced by some worst-case learning process. Denote this learning process by $\Pi_1, \Pi_{2,1}$. 

\item Nature then provides $\pi_1$ according to $\Pi_{1}$ from the previous step to the buyer. 

\item At $t=2$, the seller maximizes profit assuming the worst-case information structure \emph{at $t=2$}, holding fixed $\pi_1$. Denote this information structure by $\pi_{2,2}$. This determines the equilibrium price $p_2$. 

\item In particular, we require $p_2$ to be consistent with $\sigma_2(h_S^2)$, while $\pi_{2,2}$ is not necessarily consistent with $\Pi_{2,1}$.
\end{itemize}

This model is substantially more complicated than the benchmark, since it requires multiple fixed-point arguments and the existence of such a profile is not obvious. This alternative benchmark provides a new interpretation of Theorem \ref{thm:selfconfirming}: under threshold-ratio monotonicity, the price path chosen by a sophisticated maxmin seller coincides with the price path in the main model. The reason is straightforward: under threshold-ratio monotonicity, the one-shot worst case always coincides with the sequentially worst case in each period along the equilibrium path. Thus, under threshold-ratio monotonicity, the price path in the main model can be viewed as a special case of the sophisticated maxmin seller (special in the sense that $\pi_{2,2}$ is consistent with $\Pi_{2,1}$).

In general, however, the sophisticated benchmark differs from the one in this model. We present an example in Section \ref{sect:sophistexample}—featuring discrete values—where the equilibrium learning process is not the one required to induce the outcome described in Theorem \ref{thm:baseline}.\footnote{The assumption of discrete values does not change the analysis relative to the continuous case; we discuss why continuous distributions that approximate discrete ones will typically violate threshold-ratio monotonicity.} Beyond this, we are not able to say much more. Solving for the equilibrium price paths under this alternative, even in simple examples, is beyond the scope of existing techniques we are aware of, and thus for now we leave it as an open problem.\footnote{For instance, the approach of \cite{ACM2022}, who derive an HJB representation for a sophisticated maxmin decision maker, does not directly apply in our setting, since it is not clear which state variable one could use. The natural choice (and the choice in \cite{ACM2022}) would be the set over which the seller has uncertainty at time $t$. However, the set of possible Nature choices from time $t$ onward does not pin down the seller's payoff, since past information structures influence which buyers have already purchased or remain in the market, and thus matter for the seller's continuation value. Note that in \cite{ACM2022}, Nature's choice at time $t$ is the \emph{initial} prior, making their setting closer to Section \ref{sect:worsepast} than Section \ref{sect:sophistication}.} While we expect the resulting price paths to be qualitatively similar, the key point for our purposes is the following: the resulting equilibrium can be interpreted as displaying non-Coasian forces, since both our model and this alternative induce identical single-period problems but different dynamic solutions.

\subsubsection{Example of Sophsticated Maxmin Differing from Sequentially Worst Case} \label{sect:sophistexample}
Consider the discrete distribution where $v=1$ with probability $1/2$ and $v=0$ with complementary probability. The concavification arguments of \cite{KG2011} immediately imply that the worst case makes the buyer indifferent between purchasing and not whenever recommended to not purchase. Therefore, in the static problem, given a price $p <1/2$, the information structure recommends purchase with probability $r$ when $v=1$, where $r$ satisfies
\begin{equation*}
p = \frac{(1-r)q}{(1-r)q + 1 - q} 
\quad \Rightarrow \quad 
r = \frac{q - p}{q(1-p)},
\end{equation*}
where $q$ is the prior that $v=1$.  

Although our model assumed a continuous value distribution, this is not essential to obtain Theorem \ref{thm:baseline} in the two-period case. In the second period, Nature induces expectation $p_{2}$, which generates no additional option value, and in the first period, Nature induces expectation $w(p_{1})$, the indifference value for a consumer facing price $p_{1}$. Given $w(p_{1})$, the second-period price maximizes
\begin{equation*} 
p_{2}\left( \frac{w(p_{1}) - p_{2}}{w(p_{1})(1-p_{2})} \right),
\end{equation*}
since $w(p_{1})$ is also the probability that $v=1$ in period 2. Maximizing this over $p_{2}$ yields 
\[
p_{2} = 1 - \sqrt{1-w(p_{1})}.
\]
Using this, we can solve for $w(p_{1})$ from the indifference condition
\[
w(p_{1}) - p_{1} = \delta \big(w(p_{1}) - p_{2}\big).
\]
Given a solution for $w(p_{1})$ (assuming it is interior), $p_{1}$ is then chosen to maximize
\begin{equation*} 
\frac{1}{2} \cdot p_{1}\left(\frac{1/2 - w(p_{1})}{(1/2)(1-w(p_{1}))}\right)
+ \left(\tfrac{1}{2} + \tfrac{1}{2}\left(1-\frac{1/2 - w(p_{1})}{(1/2)(1-w(p_{1}))}\right)\right)
\cdot w(p_{1}) \delta p_{2}(p_{1})\left(\frac{w(p_{1}) - p_{2}(p_{1})}{w(p_{1})(1-p_{2}(p_{1}))}\right).
\end{equation*}
This expression can be maximized numerically; doing so for $\delta = 2/3$ yields the following solution: 
\begin{equation*} 
p_{1} \approx 0.2609, \quad  w(p_{1}) \approx 0.3700, \quad  p_{2} \approx 0.2072, \quad \text{Seller Payoff} \approx 0.0763.
\end{equation*}

\noindent For this price path, we verify that the resulting solution is not safe, and hence that the sophisticated fully maxmin seller would use a different pricing strategy than the one outlined above. Suppose instead that the seller charged prices $p_{1}$ and $p_{2}$ as above, but Nature used an information structure that perfectly revealed the value to the buyer in the second period. In this case, the buyer would optimally delay, since when $\delta = 2/3$: 
\begin{equation*} 
\tfrac{1}{2} - 0.2609 \;<\; \tfrac{1}{2}\cdot \tfrac{2}{3}(1 - 0.2072).
\end{equation*}

\noindent On the other hand, the seller's payoff under this alternative---where the buyer purchases at time 2 whenever $v=1$---is $(2/3)(1/2)p_{2} \approx 0.0691 < 0.0763$. Thus, the fully worst-case learning process is not the one previously identified. 

While threshold-ratio monotonicity is only defined for continuous distributions, we note that it will be violated for continuous distributions approximating this discrete case—for instance, by taking $n$ even and sufficiently large and considering 
\[
f(v) = (v - 1/2)^{n}(1+n)2^{n}.
\]
Intuitively, for moderate values of $v$—say, in the range $[1/4,1/3]$—and for $n$ very large, the threshold $y^{*}(v)$ will be very close to $1$ for all values in this range. As a result, over this interval, $y(v)$ increases only slightly as $v$ increases, even with large changes in $v$. Hence, the ratio $\tfrac{v}{y(v)}$ will increase as well.

\subsection{Worse Past Information} \label{sect:worsepast}
We have assumed that the seller treats all \emph{past} actions of Nature as ``sunk.'' Since the seller knows Nature has already moved, a seller setting a price at time $t$ does not consider the worst-case information structure at any $s < t$—that is, these information structures are taken as known. However, if the seller at time $t$ were to consider the worst case over \emph{all} information arrival processes, these could include past information as well. 

Specifically, assume the following, and for simplicity\footnote{While there is no conceptual difficulty in considering the general time-horizon case, doing so formally requires additional technical details regarding the definition of equilibrium.} take $T=2$: 

\begin{itemize} 
\item At time 1, the timing protocol is exactly as in the main model. 

\item At time 2, the seller chooses a price to maximize the profit guarantee, taken over all $\pi_1, \pi_2$, conditional on the buyer not having purchased at time 1. 
\end{itemize}

\noindent To obtain a coherent statement while avoiding conceptual difficulties, we treat the buyer as a completely passive player and do not consider their incentives, taking $\hat{p}_{2}(p_{1})$ as primitive—a more complete model would require an assumption about how this is determined. 

For this model, Section \ref{app:proofB3} presents a result showing that in the two-period setting, and under the restrictive assumption that sales occur in both periods with positive probability, the worst-case ``past information structure'' from the perspective of time 2 is one where the time-1 information was \emph{most favorable} to the seller. This corresponds to the buyer being informed whether $v$ lies above or below a threshold, but where the buyer is indifferent between buying and not \emph{whenever buying} (as opposed to whenever delaying). 

The intuition for this result is that Nature can condition on the fact that the buyer has not purchased when choosing a ``past information structure.'' This is still restricted, since coherence requires that the information structure be such that the buyer would have been willing to purchase under the conjecture. But by selecting past information in this way, Nature ensures that the buyers who remain have the lowest possible values.

We do not present a full characterization of equilibrium for this benchmark for two reasons. First, doing so formally requires specifying how the seller resolves his time inconsistency, as well as how the seller believes the buyer believes the seller resolves his time inconsistency. At $t=1$, the problem appears to the seller exactly as in the model described in Section \ref{sect:model}, but at $t=2$ the problem appears very different. Thus, there are (at least) two possible candidates for $\hat{p}_{2}(p_{1})$, and without an assumption on (the seller’s belief of) buyer equilibrium behavior, we cannot specify which first-period indifference threshold is relevant.

Second, characterizing the full equilibrium requires identifying primitive conditions that ensure sale occurs in both periods with positive probability, in order to avoid making assumptions on endogenous objects. Without this assumption, the seller could form a $t=2$ conjecture implying that the buyer should have purchased at $t=1$ with probability one. If this were possible, the seller would then believe himself to be at a probability-zero event \emph{whenever} the game continues to time 2. We wish to avoid taking a stand on how the seller disciplines beliefs in this case.  

Still, this discussion clarifies the kind of dynamic inconsistency issues that arise when the seller allows the worst case to extend to past information. The result strikingly suggests that the seller always believes the past information was chosen favorably, despite future information being unfavorable. We leave our analysis of this alternative at this observation.

\subsubsection{Proof of the above claim} \label{app:proofB3}

We now present a formal statement of the result alluded to in the previous section:

\begin{proposition}  \label{prop:alsopast}
Suppose $T=2$, and that at time 2 the seller seeks to maximize the profit guarantee over the worst-case choices of 
\[
\tilde{\pi}_{1} : [\underline{v}, \overline{v}] \to \Delta(S_{1}) 
\quad \text{and} \quad 
\pi_2 : [\underline{v}, \overline{v}] \times S_{1} \to \Delta(S_{2}).
\] 
Suppose further that, at time 2, the seller conjectures that the buyer anticipated a second-period price $\hat{p}_{2}(p_{1})$. Let 
\[
v^{*} = \frac{p_{1} - \delta \hat{p}_{2}(p_{1})}{1 - \delta},
\] 
and assume $v^{*} > \E_{F}[v]$. Then, given a price $p_{1}$, the worst-case $\tilde{\pi}_{1}$ (for the seller at time 2) involves the buyer learning whether $v > y^{*}$, where $y^{*}$ is either equal to $\underline{v}$ or characterized by 
\[
\E[v \mid v > y^{*}] = \frac{p_{1} - \delta \hat{p}_{2}(p_{1})}{1 - \delta}.
\]
\end{proposition}

\begin{proof}[Proof of Proposition \ref{prop:alsopast}] We consider the time-1 problem, and assume without loss assume all signals are collapsed to action recommendations via a revelation argument. Letting $\overline{s}_{1}$ denote the time-1 recommendation for the buyer to buy and $\underline{s}_{1}$ denote the time-1 recommendation to not buy, note that we cannot have $\mathbb{E}[v \mid \overline{s}_{1}] <  \frac{p_{1} - \delta \hat{p}_{2}(p_{1})}{1 - \delta}$,  since otherwise the information structure would not be obedient and such buyers would rather not buy.   On the other hand, suppose $\mathbb{E}[v \mid \overline{s}_{1}] > \frac{p_{1} - \delta \hat{p}_{2}(p_{1})}{1 - \delta}$.  Consider an alternative information structure where,  whenever $\underline{s}_{1}$ is drawn,  the recommendation is switched to $\overline{s}_{1}$ with probability $\varepsilon$ (and otherwise remains the same).  Since this modification does not change the conditional distribution but scales the total mass by $1-\varepsilon$, the seller's optimal second period price is unchanged.  But since the total mass of buyers remaining at time 2 under this modification decreases, this hurts the seller. 

It follows that $\mathbb{E}[v \mid \overline{s}_{1}] =  \frac{p_{1} - \delta \hat{p}_{2}(p_{1})}{1 - \delta}$ must hold.   Consider any information structure that is not partitional with this property.   Let $\bar{u}(v)$ denote the probability that the buyer is recommended to  buy at time 1 with value $v$;  for $u \sim U[0,1]$,   we can without loss consider an implementation where the information structure recommends ``buy'' whenever $u > \bar{u}(v)$.  Note that an information structure is partitional if and only if $\bar{u}(v)$ is a step function (outside of a set of measure 0). 

For $\varepsilon$ small,  let $\tilde{v}(\tilde{\varepsilon})$ be such that $\int_{\underline{v}}^{\tilde{v}(\varepsilon)} \bar{u}(v)f(v) = \varepsilon$.  Consider a modification where (i) all buyers with $v < \tilde{v}(\tilde{\varepsilon})$ are recommended to not purchase at time 1,   and (ii) all buyers with  $v > \hat{v}(\varepsilon)$  purchase at time 1,  with $\hat{v}(\varepsilon)$ chosen so that $\mathbb{E}[v \mid \overline{s}_{1}] =  \frac{p_{1} - \delta \hat{p}_{2}(p_{1})}{1 - \delta}$ still holds given the modification (i).   First, note there exists $\varepsilon$ sufficiently small so that  $\tilde{v}(\varepsilon) < \hat{v}(\varepsilon)$ if and only if $\bar{u}(v)$ is not a step function, since the information structure is not partitional if and only if the minimum of the support of $v$ given $\bar{s}$ is strictly less than the maximum of the support of $v$ given $\underline{s}$, and since these are $\lim_{ \varepsilon \rightarrow 0} \hat{v}(\varepsilon)$ and  $\lim_{ \varepsilon \rightarrow 0} \tilde{v}(\varepsilon)$, respectively.  Since $\mathbb{E}[v \mid \overline{s}]$ is unchanged,  but with the lowest values no longer receiving $\overline{s}$,  we must have $\int_{\underline{v}}^{\tilde{v}(\varepsilon)} \bar{u}(v) f(v) dv <  \int_{\hat{v}(\varepsilon)}^{\overline{v}} \bar{u}(v) f(v) dv$.  

Now,  for $\varepsilon$ sufficiently small, remaining buyers with $v > \hat{v}(\varepsilon)$ buy with probability 1 in the second period under the original information structure (but never in the replacement),  since $\hat{v}(\varepsilon)$ converges to the maximum of the support of time-2 values.  But the increase in the sale probability is at most $\int_{\underline{v}}^{\tilde{v}(\varepsilon)} \bar{u}(v) f(v) dv$. Since $\int_{\underline{v}}^{\tilde{v}(\varepsilon)} \bar{u}(v) f(v) dv <  \int_{\hat{v}(\varepsilon)}^{\overline{v}} \bar{u}(v) f(v) dv$, this modification hurts the seller. So this modification must be infeasible, making the information structure partitional. \end{proof}

\end{document}